\documentclass[onecolumn]{els-mrw} 

\usepackage{amsmath,amssymb,amsfonts,amsthm,makeidx,graphicx}
\usepackage{amsmath,dsfont,mathtools,amsfonts,bbold,physics}
\graphicspath{{./img/}}

\usepackage{txfonts}
\usepackage{helvet}

\begin{document}

\chapter{Quantum Kicked Top: A Paradigmatic Model}\label{chap1}

\author[1]{Avadhut V. Purohit}%
\author[1]{Udaysinh T. Bhosale}%


\address[1]{\orgname{Visvesvaraya National Institute of Technology}, \orgdiv{Department of Physics}, \orgaddress{Nagpur 440010, India}}


\maketitle

\begin{glossary}[Keywords]
Classical and quantum chaos, Floquet system, random-matrix theory, entanglement

\end{glossary}

\begin{abstract}[Abstract]
    The quantum kicked top (QKT) is one of the most widely studied models in quantum chaos, providing a minimal yet powerful framework for exploring the relationship between classical nonlinear dynamics and quantum behavior. Unlike many chaotic systems with infinite-dimensional Hilbert spaces, the QKT possesses a finite-dimensional Hilbert space, making it analytically and numerically controllable while still showing rich dynamical phenomena. In this chapter, we present a comprehensive introduction to the QKT as a paradigmatic model of quantum chaos. Starting from the classical kicked top, we derive the discrete nonlinear map governing the dynamics on the unit sphere and analyze its phase space structure through fixed points, stability analysis, bifurcations and Lyapunov exponents. We then discuss the role of symmetries, including rotational and time-reversal symmetry, and how their breaking modifies the dynamics. The quantum description is developed using Floquet theory, where the periodically driven spin system is represented by a unitary Floquet operator acting on a $(2j+1)$-dimensional Hilbert space. Within this framework, signatures of quantum chaos such as spectral statistics, entanglement generation and recurrences are discussed. The model also admits an interpretation as a system of interacting qubits, enabling explicit few-qubit realizations and direct connections with quantum information measures through reduced density matrices and entanglement entropy. By linking classical phase space structures with quantum dynamical indicators, the QKT provides a clear setting to investigate the emergence of chaotic behavior in the semiclassical limit. The chapter, therefore, highlights the quantum kicked top as a bridge between nonlinear classical dynamics, quantum chaos and modern quantum information science.
\end{abstract}

\section{Introduction}\label{sec:introduction}
Most of the systems we observe in the world are nonlinear, including the simple pendulum. If we attach another pendulum to the end of the first, we observe motion that is highly sensitive to the initial conditions, practically making it unpredictable. The motion of this double pendulum is chaotic despite being deterministic. If we quantize it, the situation becomes even more challenging. The already unpredictable trajectories will be replaced by probabilistic predictions. Macroscopic systems are made of quantum particles. Therefore, understanding the quantum principles responsible for chaotic behavior in the classical limit is necessary. Understanding quantum chaos is crucial for technologies operating from the deep-quantum to the semiclassical regime.

One can study models such as the quantum double pendulum, but the infinite-dimensional
Hilbert space makes its analysis difficult. On the other hand, the QKT~\cite{haake1987classical} is also a nonlinear system but with finite dimensional Hilbert space. It consists of a freely precessing spin particle with periodically applied nonlinear kicks. Furthermore, despite its simple construction, the QKT reveals rich and complex dynamics. This makes it an ideal tool to explore how quantum dynamics reproduce signatures of classical chaotic motion in the semiclassical limit.

Researchers have shown that different statistical methods can produce distinct results in infinite-dimensional Hilbert spaces~\cite{haake1987classical, casati1980connection,Izrailev1989}. As Casati et al. have discussed an equivalence of $\mathcal{U}$ to certain ensembles of random matrices is much more difficult to establish for an infinite dimensional Hilbert spaces~\cite{casati1980connection}. Izrailev have also pointed out that the eigenvalues of the Floquet operator for the kicked rotor gives drastically different statistical properties if we use different truncation methods~\cite{Izrailev1989}. The finite dimensionality of the QKT avoids these ambiguities. Additionally, it allows clear exploration of the entire regime from the deep quantum to semi-classical behavior. The model provides an ideal platform for investigating ideas, including the application of the random matrix theory (RMT), the study of out-of-time-order correlators (OTOCs), quantization around classically short-lived regular structures, and the analysis of entanglement dynamics. This makes the QKT a paradigmatic model.

Here, the term `quantum chaos' is often used loosely, and there is no exponential sensitivity to initial conditions in the quantum domain. Researchers have different approaches to study quantum chaos. Some look for the quantum version of the Lyapunov exponent, others argue for entirely new quantities such as entanglement to characterize chaos. The QKT, with its finite-dimensional Hilbert space and tunable nonlinear dynamics, provides a platform to address these issues. The model, with its variations and extensions, therefore, plays a significant role in expanding the literature of quantum chaos.

As a spin system with all-to-all nonlinear interactions, the QKT provides fertile ground to study quantum correlations such as entanglement, quantum discord, and concurrence. Unexpectedly, these quantum correlations have been shown to mimic the classical Lyapunov exponent. This model has been experimentally realized in few-qubit systems, enabling precise measurements of entanglement dynamics. The model has been used in quantum control strategies, enhancing fidelity in quantum gates and simulation of noise in quantum computing. These studies make the case for the QKT as a practical model in the emerging quantum technologies.

This chapter aims to present the QKT as a finite-dimensional model of chaos that bridges classical nonlinear dynamics and quantum behavior. It uses this single framework to support rigorous analysis through maps, fixed points, Lyapunov exponents, and symmetries, along with quantum-chaos signatures such as entanglement, spacing statistics, and recurrences. The goal is to show how changing the kick strength drives the system from regular motion to global chaos. It also reveals how classical structures leave measurable traces in few-qubit quantum systems and in the semiclassical regime.

Even though the reader is not expected to be familiar with the following concepts and mathematical techniques, their knowledge would make life easier:
\begin{itemize}
    \item Classical Hamiltonian dynamics and phase space concepts: fixed points, stability analysis, bifurcations, Lyapunov exponents, and maps on a constrained surface (e.g., motion on a sphere).
    \item Basic group-theoretic quantum mechanics: angular momentum operators \(J_x, J_y, J_z\), their commutation relations, and the spin-\(j\) representation acting on a \((2j+1)\)-dimensional Hilbert space.
    \item Floquet theory: evolution over one period, Floquet operators, and their dynamics.
    \item Introductory quantum information theory: qubits as spin-\(1/2\) systems, multi-qubit tensor products, reduced density matrices (RDMs), and entanglement measures.
\end{itemize}

By the end of the chapter, the reader would acquire:
\begin{itemize}
    \item A detailed study of the QKT: the derivation of the classical map, and the role of parameters \((k,p)\) in governing regular and chaotic dynamics.
    \item Tools in the classical dynamics: fixed points and their stability analysis, bifurcation cascades (period-doubling), and the use of the largest Lyapunov exponent (LLE) as an effective chaos indicator.
    \item An understanding of symmetries: time-reversal symmetry and its breaking, the role of rotational symmetry in governing the dynamics.
    \item A bridge to quantum dynamics: interpretation of the same model as a system of few interacting qubits, explicit solutions for 2- to 4-qubit systems, RDMs, and infinite-time-averaged linear entropy landscapes that mimic classical phase space structures.
    \item An understanding of eigenvalue and eigenvector statistics and of dynamical signatures, such as the Loschmidt echo and spectral form factor, as tools for detecting quantum chaos.
    \item Implementations of the QKT in various experimental setups.
\end{itemize}
This way, the chapter prepares the reader for advanced topics in quantum chaos, quantum nonlinear systems and quantum information.

\section{Classical Dynamics}
The work of Boris Chirikov on the resonance-overlap criterion and later the standard (Chirikov) map laid the foundation for models of Hamiltonian chaos such as the kicked rotor~\cite{chirikov1979universal}. Fritz Haake and collaborators extended these ideas to the quantum spin with finite-dimensional Hilbert spaces~\cite{haake1987classical}. This construction avoids ambiguities (discussed in the previous Sec.~\ref{sec:introduction}) associated with the infinite dimensionality of the Hilbert space of the quantum kicked rotor, while maintaining rich interplay between order and chaos. In this model, the rotor is replaced by a top that undergoes continuous precession about the $y$-axis and is subjected to periodic nonlinear kicks around the $z$-axis. The Hamiltonian operator of the model is given by
\begin{align}
    H = \frac{\hbar p}{\tau} J_y + \frac{\hbar k}{2j} J_z^2 \sum_{n=-\infty}^{\infty} \delta \left(t - n \tau \right).
\end{align}
The operators $J_x, J_y, J_z$ are angular momentum operators acting on a $(2j + 1)$-dimensional Hilbert space; $p/\tau$ denotes the angular frequency and $k$ is the kicking strength. Then, the Floquet operator $\mathcal{U}$ is given as follows:
\begin{align}
    \mathcal{U} = \mathcal{U}_k \mathcal{U}_p = \exp\left(-i\frac{k}{2j}J_z^2\right) \exp\left(-i p J_y\right).
\end{align}

\subsection{Classical Map}\label{sec:sub:classical_map}
The classical Hamiltonian dynamics requires us to find the classical map corresponding to the above quantum evolution. This can be obtained by evolving the angular momentum vector $\mathbf{J} = (J_x, J_y, J_z)$:
\begin{align}
    \mathbf{J}' = \exp\left(i p J_y\right) \exp\left(i \frac{k}{2j}J_z^2\right) \mathbf{J} \exp\left(- i \frac{k}{2j}J_z^2\right) \exp\left(- i p J_y\right),
\end{align}
and then taking the classical limit, $\mathbf{X} = \lim_{j\to\infty} \frac{\mathbf{J}}{j}$. This map is given by (see Appendix~\ref{appendix:classical_map} for detailed derivation)
\begin{align}\label{Eq:classical_map}
    X' &= \cos \left[k(-X \sin p + Z \cos p)\right] (X \cos p + Z \sin p) - \sin \left[k(-X \sin p + Z \cos p)\right] Y, \notag \\
    Y' &= \cos \left[k(-X \sin p + Z \cos p)\right] Y + \sin \left[k(-X \sin p + Z \cos p)\right] (X \cos p + Z \sin p), \notag \\
    Z' &= -X \sin p + Z \cos p,
\end{align}
satisfying constraint $X^2 + Y^2 + Z^2 = 1$. The classical dynamics of the kicked top are determined by two parameters: the kicking strength $k$ and the precession angle $p$. For a particular point $\mathbf{X}_n$ on the phase space, the above classical map gives the next point $\mathbf{X}_{n+1}$. The trajectory of the kicked top is then obtained by iterating this map over discrete time $n \in \mathds{Z}$.

\subsection{Fixed Point Analysis}
Fixed points are specific points in the phase space where the trajectory returns to the same point after every iteration. Mathematically, a (period-1) fixed point $\mathbf{X}^*$ satisfies the condition $\mathbf{X}^* = \mathbf{F}(\mathbf{X}^*)$. Usually, trajectories in the neighbourhood of stable fixed points are fixed orbits. This idea is generalized to the period-$n$ fixed points, where the trajectory returns to the initial point after $n$ iterations. The orbits surrounding the stable ones are called period-$n$ orbits. Therefore, fixed points and their stability hold a key in understanding the overall dynamics of the system.

The point $\mathbf{X}$ is called period-$n$ fixed point if it satisfies the following condition:
\begin{align}
    \mathbf{F}^n(\mathbf{X}) = \mathbf{X} \quad\text{and}\quad \mathbf{F}^m(\mathbf{X}) \neq \mathbf{X} \quad \forall \; m < n.
\end{align}
Here, $\mathbf{F}^n(\mathbf{X}) = \mathbf{F}( \mathbf{F}^{n-1}(\mathbf{X}) )$ is an iterative classical map. The stability of these fixed points is determined by the Jacobian matrix $\mathbf{M}_{3\times 3}$ of the classical map $\mathbf{F}$ evaluated at a particular point. It is given by
\begin{align}
    \mathbf{M} = \frac{\partial \mathbf{F}}{\partial \mathbf{X}} \bigg|_{\mathbf{X} = \mathbf{X}^*}.
\end{align}
Even though the system is three-dimensional, the constraint $X^2 + Y^2 + Z^2 = 1$ reduces the effective dimensionality to two. This makes one eigenvalue among three always equal to unity and simplifies the characteristic equation to a quadratic form:
\begin{align}
    \left(\lambda-1\right) \left[\lambda^2 + (1-\text{Tr}(\mathbf{M})) \lambda + \det(\mathbf{M}) \right] = 0.
\end{align}
The given fixed point $\mathbf{X}^*$ is stable if both eigenvalues of the Jacobian matrix $\mathbf{M}$ lie on the unit circle in the complex plane. That is, 
\begin{align}
    \bigg| \frac{\text{Tr}(\mathbf{M}) - 1 \pm \sqrt{{\left(\text{Tr}(\mathbf{M}) - 1\right)}^2 - 4}}{2}  \bigg| < 1 \implies \big| \text{Tr}(\mathbf{M}) - 1\big| < 2.
\end{align}

The fixed points corresponding to the classical map are obtained by setting $\mathbf{F}\left(\mathbf{X}\right) = \mathbf{X}$. Solving these equations leads to the following set of fixed points for the map~\eqref{Eq:classical_map}:
\begin{align}
    Z &= - X \cot \left(\frac{p}{2}\right), \quad
    Y = X \cot \left[\frac{k X}{2} \cot\left(\frac{p}{2}\right)\right] \\
    f\left(X\right) &= \frac{\sin^2\left[\frac{k X}{2} \cot\left(\frac{p}{2}\right)\right]}{1 + \cot^2\left(\frac{p}{2}\right) \sin^2\left[\frac{k X}{2} \cot\left(\frac{p}{2}\right)\right]} - X^2 = 0.
\end{align}
It can be seen that $\mathbf{X}=(0, \pm 1, 0)$ are trivial fixed points. When $p = 0$, nontrivial fixed points do not arise with $k$. However, for $p = \pi/2$, the situation becomes interesting. The fixed points arise and become unstable for different values of $k$. The classical map for this case gets simplified as follows:
\begin{align}
    \mathbf{F} = \begin{pmatrix}
        Z \cos(k X) + Y \sin(k X) \\
        -Z \sin(k X) + Y \cos(k X) \\
        -X
    \end{pmatrix}.
\end{align}
The above-mentioned period-1 fixed points also get simplified as follows:
\begin{align}\label{Eq:fixedPoint:period1}
    Z = - X, \quad
    Y = X \cot \left(\frac{k X}{2}\right) \quad \text{and} \quad
    f\left(X\right) = \frac{\sin^2\left(\frac{k X}{2}\right)}{1 + \sin^2\left(\frac{k X}{2}\right)} - X^2 = 0.
\end{align}
Since the Jacobian matrix $\mathbf{M}$ at this point has $\text{Tr}(\mathbf{M}) = kX \cot\left(kX/2\right) + \cos\left(kX\right)$, its stability is given by
\begin{align}
    \bigg| kX \cot\left(\frac{kX}{2}\right) + \cos\left(kX\right) - 1\bigg| < 2.
\end{align}
For small values of $k$, it can be seen that
\begin{align}
    f(X) \approx {\left(\frac{k X}{2}\right)}^2 - X^2 = 0 \implies X = 0 \quad \text{or} \quad k = 2.
\end{align}
Thus, for $0 < k < 2$, trivial fixed points are stable, as illustrated in Fig.~\ref{fig:pahse_portrait}(a). At $k = 2$, they lose stability and bifurcate. The bifurcation of the point $(0, 1, 0)$ creates two new period-1 fixed points. Whereas, a bifurcation of the point $(0, -1, 0)$ creates two new period-2 points. This distinction between positive and negative $\phi$ domains arises due to the symmetry of our classical map. It will be discussed later in the subsection devoted to symmetries. For new period-1 fixed points arising at $k>2$, the condition $f(X)=0$ cannot be satisfied beyond $X=1/\sqrt{2}$. These fixed points lose their stability at $k = \sqrt{2}\pi$. Beyond this value of $k$, each fixed point bifurcates into new period-2 fixed points. The period-2 fixed points on this symmetry line $Z=-X$ are given by:
\begin{align}\label{Eq:fixedPoint:period2}
    Z_1 = Z_2 = - X, \quad Y_1 = -Y_2 = \sqrt{1-2X^2} \quad \text{and} \quad X = (2m+1)\frac{\pi}{k}.
\end{align}
The stability of these period-2 fixed points is determined by the Jacobian matrix $\mathbf{M}$ evaluated at $\mathbf{F}^{(2)}(\mathbf{X}) = \mathbf{X}$. The stability condition is, thus, given by
\begin{align}
    \bigg| 2 - k^2 + 2\pi^2 {\left(2m + 1\right)}^2  \bigg| < 2 \implies \sqrt{2}\pi \left(2m + 1\right) < k < \sqrt{4 + 2\pi^2 {\left(2m + 1\right)}^2}.
\end{align}
Beyond this value of $k$, these fixed points lose their stability and bifurcate into new higher-period fixed points. This process continues, leading to the creation of higher-period fixed points through bifurcations as $k$ increases.

Other than these, there are fixed points such as $(\pm 1, 0, 0)$ and $(0, 0, \pm 1)$ that do not lie on the symmetry line $Z = -X$. The stability of these period-4 fixed points is given by
\begin{align}
    {\left(2\cos^2(k) + k \sin(k)\right)}^2 < 4.
\end{align}
Similar to the earlier cases, these fixed points also lose their stability and go on to create cascaded higher-period fixed points through bifurcations as $k$ increases.

\subsection{Symmetries}
Symmetries of the kicked top constrain the classical phase space and organize phase space structures into symmetry-related families. In particular, the rotational and time-reversal symmetries considered below provide a direct connection between symmetry-related periodic orbits and their stability properties. We, therefore, analyze these symmetries before discussing their consequences for the periodic orbits and phase space structure. The classical map $\mathbf{F}$ is observed to be invariant under rotation around the precession axis by an angle $\pi$. This rotational symmetry can be expressed as follows:
\begin{align}\label{Eq:Ry:symmetry}
    R_y(\pi) \begin{pmatrix}
        X \\
        Y \\
        Z
    \end{pmatrix} = \begin{pmatrix}
        -X \\
        Y \\
        -Z
    \end{pmatrix} \implies \mathbf{F}(R_y(\pi) \mathbf{X}) = R_y(\pi) \mathbf{F}(\mathbf{X}).
\end{align}
It shows that the $R_y$-image of every period-$n$ orbit is also a period-$n$ orbit. 

Even though the classical map does not have symmetry under rotation around the $x$-axis by an angle $\pi$, when this rotation is combine dwith the $R_y$ symmetry, the map satisfies the following property:
\begin{align}\label{Eq:Rx:operator}
    R_x(\pi) \begin{pmatrix}
        X \\
        Y \\
        Z
    \end{pmatrix} = \begin{pmatrix}
        X \\
        -Y \\
        -Z
    \end{pmatrix} \implies \mathbf{F}(R_x(\pi) \mathbf{X}) = R_x(\pi) \mathbf{F}(R_y(\pi) \mathbf{X}).
\end{align}
Applying the classical map once more to both sides of Eq.~\eqref{Eq:Rx:operator} and using Eq.~\eqref{Eq:Ry:symmetry}, we get
\begin{align}\label{Eq:Rx:symmetry}
    \mathbf{F}^{(2)}(R_x(\pi) \mathbf{X}) = R_x(\pi) \mathbf{F}^{(2)}(\mathbf{X}).
\end{align}
This shows that every period-$n$ orbit (for odd-$n$) together with its $R_y$-image, gets mapped by $R_x$ into a period-$2n$ orbit of $\mathbf{F}$.

The kicked top exhibits a non-conventional time-reversal symmetry. For a given Floquet operator $\mathcal{U}$, if there exists an operator $\mathcal{T}$ such that
\begin{align}
    T \mathcal{U} T^{-1} = \mathcal{U}^\dagger,
\end{align}
then the system is said to have time-reversal symmetry. Using this property, we can check that the operator given by
\begin{align}
    T = \exp\left(i p J_y\right) \exp\left(i \pi J_z\right) K,
\end{align}
acts as a time-reversal operator for the kicked top. For a detailed derivation, see the supplementary material of Ref.~\cite{purohit2025double}. Here, $K$ is the complex conjugation operator. Due to $R_y(\pi)$-symmetry given by Eq.~\eqref{Eq:Ry:symmetry}, the kicked top also exhibits another inequivalent time-reversal operator given by
\begin{align}
    \tilde{T} = \exp\left(-i \pi J_x\right) \exp\left(-i p J_y\right) K.
\end{align} 
In the classical limit, these time-reversal symmetries translate to the following properties of the classical map $\mathbf{F}$:
\begin{align}
    T\begin{pmatrix}
        X \\
        Y \\
        Z
    \end{pmatrix} = \begin{pmatrix}
        -X \cos(p) - Z \sin(p) \\
        Y \\
        Z \cos(p) - X \sin(p)
    \end{pmatrix} \quad \text{and} \quad
    \tilde{T}\begin{pmatrix}
        X \\
        Y \\
        Z
    \end{pmatrix} = \begin{pmatrix}
        X \cos(p) + Z \sin(p) \\
        Y \\
        -Z \cos(p) + X \sin(p)
    \end{pmatrix}.
\end{align}

The fixed-point analysis in the preceding subsection concerns the $T$-symmetric family. However, a similar analysis can be done for $\tilde{T}$-symmetry as well. The period-2 fixed points for this symmetry are given by
\begin{align}
    Z_1 = X, \quad Y_1 = -X \cot \left(\frac{k X}{2}\right) \quad \text{and} \quad 
    Z_2 = -X, \quad Y_2 = -X \cot \left(\frac{k X}{2}\right).
\end{align}
Here, the period-1 fixed point given by Eq.~\eqref{Eq:fixedPoint:period1} together with its $R_y$-image gets mapped by $R_x$ into a period-2 fixed point of $\mathbf{F}$. Due to this symmetry, the stability condition of these period-2 fixed points is the same as that of the earlier period-1 fixed points. Similarly, the period-2 fixed points given by Eq.~\eqref{Eq:fixedPoint:period2} together with their $R_y$-images get mapped by $R_x$ into a period-4 fixed point of $\mathbf{F}$. The period-4 fixed points for this symmetry are given by
\begin{align}
    Z_1 &= -Z_2 = Z_3 = -Z_4 = X_1, \quad Y_1 = -Y_2 = Y_3 = -Y_4 = \sqrt{1-2X_1^2} \notag \\
    X_2 &= -X_3 = X_4 = - X_1 \quad \text{and} \quad X_1 = (2m+1)\frac{\pi}{k}.
\end{align}
The stability condition for these period-4 fixed points is the same as that of the earlier period-2 fixed points. Continuing this process, one finds that odd-period points,
together with their $R_y$-images, are mapped by $R_x$ into period-$2n$ fixed points of $\mathbf{F}$.

There are interesting period-3 fixed points belonging to $T$-symmetry:
\begin{align}
    \begin{pmatrix}
        X\\
        Y\\
        -X
    \end{pmatrix}, \;
    \begin{pmatrix}
        - X \cos (kX) + Y \sin (kX)\\
        Y \cos (kX) + X \sin (kX)\\
        -X
    \end{pmatrix} \text{ and }
    \begin{pmatrix}
        X\\
        Y \cos (kX) + X \sin (kX)\\
        X \cos (kX) - Y \sin (kX)
    \end{pmatrix},
\end{align}
where $X$ and $Y$ can be obtained by solving $\mathbf{F}^{(3)}(\mathbf{X})=\mathbf{X}$ and the constraints. This leads to the following set of equations:
\begin{align}
    X \cot\left[\frac{kX}{2}\cos(kX) - \frac{kY}{2} \sin(kX)\right] = X\sin(kX) + Y \cos(kX) \quad \text{and} \quad
    2X^2 + Y^2 = 1.
\end{align}
Similarly, the period-3 fixed points belonging to $\tilde{T}$-symmetry are given by
\begin{align}
    \begin{pmatrix}
        X\\
        -X\cot(kX)\\
        X
    \end{pmatrix}, \;
    \begin{pmatrix}
        0\\
        -\dfrac{X}{\sin(kX)}\\
        -X
    \end{pmatrix} \; \text{ and }\;
    \begin{pmatrix}
        -X\\
        -\dfrac{X}{\sin(kX)}\\
        0
    \end{pmatrix}.
\end{align}
There are corresponding period-6 fixed points due to $R_x$-symmetry. 

It is important to note that there are some fixed points that require the study of global bifurcations. They involve the interaction of stable and unstable manifolds of different fixed orbits, leading to complex dynamical behaviors that cannot be captured by local analysis alone. The pair of such period-3 fixed points around the point $(0,-1,0)$ or equivalently $(\theta = \pi/2, \phi = -\pi/2)$ and their corresponding period-6 fixed points around the point $(0,1,0)$ or equivalently $(\theta = \pi/2, \phi = \pi/2)$ are illustrated in Figs.~\ref{fig:pahse_portrait}(a), \ref{fig:pahse_portrait}(b), \ref{fig:LLE}(a) and \ref{fig:LLE}(b).

\subsection{Phase Space}
\begin{figure}
    \includegraphics[width=1\textwidth]{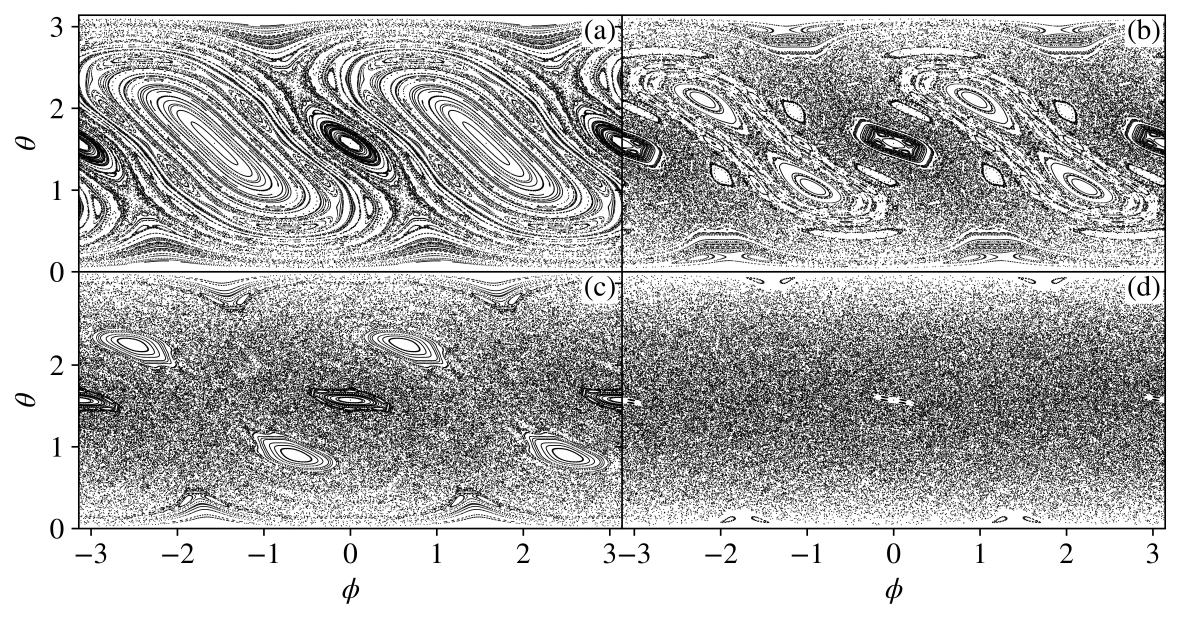}
    \caption{Classical phase space portraits for precession angle $p = \pi/2$. A grid of $25\times 25$ initial points is evolved for 200 kicks in each panel. Panels: (a) $k=2$, (b) $k=2.5$, (c) $k=3$ and (d) $k=6$.}\label{fig:pahse_portrait}
\end{figure}
The analysis of fixed points and their stability shows that the dynamics remains regular for $k < 2$. At $k=2$, the trivial fixed points lose their stability and bifurcate as shown in Fig.~\ref{fig:pahse_portrait}. Chaos starts to develop along the boundaries of period-4 orbits i.e., at $(\pm 1, 0, 0)$ and $(0,0,\pm 1)$, as illustrated in Fig.~\ref{fig:pahse_portrait}(a). In the spherical polar coordinates, these points correspond to $(\theta = \pi/2, \phi = 0)$, $(\theta = \pi/2, \phi = \pi)$, $(\theta = 0, \phi = 0)$ and $(\theta = \pi, \phi = 0)$ respectively. The rest of the phase space remains regular. At $k=2.5$, the chaotic region expands around the bifurcated trivial fixed points (see Fig.~\ref{fig:pahse_portrait}(b)). This region is further surrounded by higher-period orbits. Similarly, around period-4 fixed points, $(\pm 1, 0, 0)$ and $(0,0,\pm 1)$, more structures appear and disappear with increasing $k$. As we increase $k$ further to $k=3$, the chaotic region starts dominating the phase space, leaving only small islands of regularity around bifurcated trivial fixed points and period-4 fixed points, as shown in Fig.~\ref{fig:pahse_portrait}(c). At $k=6$, the phase space becomes largely chaotic with only tiny regular islands remaining, as illustrated in Fig.~\ref{fig:pahse_portrait}(d). Beyond $k=6$, the phase space becomes globally chaotic. In other words, no stable fixed point or periodic orbit survives beyond the kicking strength $k=6$.

It is important to note that not only local bifurcations but also global bifurcations play a crucial role in shaping the phase space structure. The local stability analysis cannot capture the emergence or decay of certain fixed points. As mentioned at the end of the preceding subsection, the existence of the pair of period-3 fixed points around $(0,-1,0)$ and their corresponding period-6 fixed points around $(0,1,0)$ at $k < 2$ demonstrates the limitations of local stability analysis (see Fig.~\ref{fig:pahse_portrait}(a)). These fixed points emerge through global bifurcations involving interactions of stable and unstable manifolds associated with different periodic orbits. A complete description of such global bifurcations requires a topological analysis of the phase space, which is beyond the scope of this chapter~\cite{kuznetsov1998elements, guckenheimer2013nonlinear}. Nevertheless, this illustrates that the kicked top provides a platform for studying global bifurcations in addition to local bifurcations.

\subsection{Largest Lyapunov Exponent}
\begin{figure}
    \includegraphics[width=1\textwidth]{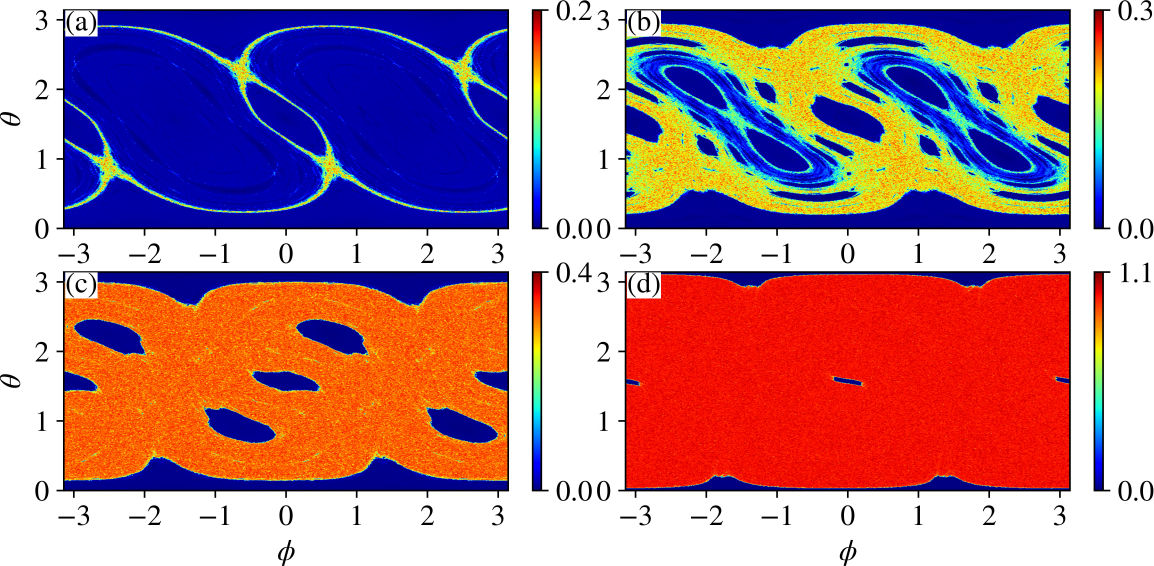}
    \caption{Largest Lyapunov Exponent for $p = \pi/2$. A grid of $500\times 500$ initial points is evolved for 1000 kicks in each panel. Panels: (a) $k=2$, (b) $k=2.5$, (c) $k=3$ and (d) $k=6$.}\label{fig:LLE}
\end{figure}
Calculating the entire spectrum of the Lyapunov exponent is computationally heavy. The largest Lyapunov exponent (LLE) is computationally less demanding, and any generic small perturbation eventually aligns with the fastest expanding direction. Therefore, the LLE serves as a sensitive, effective, and sufficient measure of chaos. A positive LLE indicates sensitive dependence on initial conditions, a hallmark of chaotic behavior, while a zero signifies regular or stable dynamics. It characterizes the average exponential rate of divergence of nearby trajectories in the phase space. It is given by
\begin{equation}
  \lambda_+ = \lim_{n \rightarrow \infty} \frac{1}{n} \ln \left[\frac{||\delta \mathbf{X}_n||}{||\delta \mathbf{X}_0||}\right].
\end{equation}
Here, $\delta \mathbf{X}_n$ is the tangent vector evolved at time $n$, which depends on the product of tangent maps. These tangent maps govern the divergence of nearby trajectories, reflecting the system's chaotic nature. The time-evolved tangent vector $\delta \mathbf{X}_n$ is given by
\begin{align}
  \delta \mathbf{X}_n &= \prod_{l=0}^{n-1} \mathbf{M}\left[\mathbf{X}_{l}\right]\cdot \delta \mathbf{X}_{0} \quad \text{where} \quad
    \mathbf{M}\left[\mathbf{X}_{l}\right] = D \mathbf{F} = \frac{\partial F_{i}}{\partial X_{j}}\Bigg|_{\mathbf{X}_{l}} \quad \text{and} \quad \\
    \prod_{l=0}^{n-1} \mathbf{M}\left[\mathbf{X}_{l}\right] &= \mathbf{M}\left[\mathbf{X}_{n-1}\right]\cdot \mathbf{M}\left[\mathbf{X}_{n-2}\right] \dots \mathbf{M}\left[\mathbf{X}_{0}\right].
\end{align}
The LLE shows excellent agreement with the stability analysis, as illustrated in Fig.~\ref{fig:LLE}. As discussed earlier, non-trivial period-$n$ orbits arising at $k=2$ can be seen in darker shades of blue (see Fig.~\ref{fig:LLE}(a)). Except for the yellow chaotic boundaries of period-4 orbits, the phase space largely remains blue, indicating regular dynamics. As kick strength increases, the chaotic region starts to expand; as a result, regular blue islands begin to shrink. The formation of a homoclinic orbit on the boundary of bifurcated trivial fixed orbits is illustrated in Fig.~\ref{fig:LLE}(b). It is again followed by the blue regular island formed by higher-period fixed orbits. At $k=3.0$, the orange chaotic sea spreads across the phase space except for those of bifurcated trivial fixed orbits and period-4 fixed orbits. This chaotic sea expands further, leaving only narrow regular islands at the trivial fixed points, and becomes more chaotic at $k=6$. Beyond this value, the phase space is fully chaotic.

\subsection{Extensions}
There are several extensions of the classical kicked top that enrich its dynamical behavior. One such extension involves generalizing two-body interactions of kicked top to $m$-body interactions~\cite{p-spin_poggi}. The Hamiltonian for such a system is given by
\begin{align}
    H(t) = \hbar \frac{p}{\tau} J_y + \hbar \frac{k}{m j^{m -1}} J_z^m \sum_{n=-\infty}^{\infty} \delta(t - n \tau).
\end{align}
The classical map for this model can be derived following a procedure similar to Sec.~\ref{sec:sub:classical_map}. Their study yields a classification of the full family of models with few-body interactions into models with $m=2$ and $m>2$. Models with $m>2$ are sub-classified into even-$m$ and odd-$m$. For $m=2$ (standard QKT), structural changes of phase space, as well as the  transition to global chaos, are dependent on both precession $p$ and kicking strength $k$. On the other hand, for $m>2$, the precession $p$ governs the structural changes in the phase space and transition to global chaos is largely governed by the kicking strength $k$ alone. Furthermore, even and odd cases of $m > 2$ are distinguished based on their bifurcations. For even $m$, some bifurcations appear in pairs due to symmetries but not for odd-$m$. This generalization extends the kicked top to an $m$-body interaction.

Another noteworthy extension is the double-kicked top. It was introduced in the seminal paper~\cite{haake1987classical} to classify quantum chaos according to symmetries. Its Floquet operator is given by
\begin{align}\label{eq:dkt}
    \mathcal{U} = e^{-i \frac{k'}{2j} J_x^2} e^{-i \frac{k}{2j} J_z^2} e^{-i p J_y}.
\end{align}
Here, kick strength $k'$ breaks the time-reversal symmetry present in the standard kicked top. A recent study analyzed this model in detail~\cite{purohit2025double}. The analysis reveals two distinct dynamics under the transformation given by
\begin{align}
    k_r = \frac{k+k'}{2} \quad \text{and} \quad k_\theta = \frac{k-k'}{2}.
\end{align}
The dynamics of $k_r$ are similar to those of the standard kicked top, leading to the transition from regularity to chaos with increasing $k_r$. The parameter $k_\theta$, on the other hand, introduces twist in phase space, resulting in the stretching of structures without affecting the overall degree of chaos. This twist parameter $k_\theta$ does not play any role in the analysis of local bifurcations. The case $k_\theta = k_r \implies (k'=0, k=2k_r)$ recovers the standard kicked top. It was observed that the broken time-reversal symmetry does bring qualitative changes in the phase space structures. However, the quantitative change in the Lyapunov exponent is negligible. This extension enables us to study the role of time-reversal symmetry or its breaking in the classical and quantum regimes.

\subsection{Lessons for Quantum systems}
The kicked top has shown us how the phase space full of regular motion at small values of kicking strength eventually becomes chaotic when the kicking strength is increased. In between these extremes, there are regular islands surrounded by chaotic seas. This transition from regularity to chaos is marked by the bifurcations of fixed points and the emergence of higher-period orbits. In addition to stable and unstable fixed points, there exist partially stable fixed points showing complex phase space. The phase space has also revealed that some bifurcations cannot be explained by local analysis involving linearization of the classical map around a certain fixed point. Furthermore, some global bifurcations have a chaotic neighborhood, but others do not. Such global bifurcations can only be explained by a topological analysis of the phase space~\cite{kuznetsov1998elements, guckenheimer2013nonlinear}. The quantization of these fixed points naturally raises questions about their quantum signatures.

The existence of a well-defined semi-classical limit motivates fundamental questions: When do the partially stable fixed points emerge? How does quantization affect their stability? Does quantum theory distinguish between local and global bifurcations? If it does, then what is their quantum origin? In the absence of a universally accepted definition of quantum chaos, the semi-classical regime allows us to do a comparative study of quantum signatures of chaos.

The kicked top, together with its nonlinear extensions, enables us to explore the interplay between the non-commutative structure of quantum mechanics and the nonlinear dynamics of the corresponding classical phase space. The quantum signatures are expected to emerge depending on the strength and form of the nonlinear interaction. Such studies then help us in addressing how these nonlinear mechanisms influence the classical-quantum correspondence in chaotic systems.

\section{Quantum Dynamics}\label{sec:quantum}
The QKT can be considered as a collection of qubits. For the total spin $j$, there are $2j$ qubits. Thus, the precession part of the Floquet operator can be given by
\begin{align}
    \exp\left(-i\frac{\pi}{2} J_y\right)  = \exp\left(-i \frac{\pi}{4} \sum_{l=1}^{2j} \sigma_l^y\right).
\end{align}
The nonlinear kicking part can be regarded as a two-body and all-to-all interaction between spin-1/2 particles. It is then given by
\begin{align}
    \exp\left(-i\frac{k}{2j} J_z^2\right) = \exp\left(-i\frac{k}{4}\right) \exp\left(-i \frac{k}{4j} \sum_{1 \leq l' < l}^{2j} \sigma_{l'}^z \sigma_l^z \right).
\end{align}
Since it can be viewed as a multi-qubit interacting system, the model is also significant in probing quantum correlations. Here, the system of a few qubits operates in the deep quantum regime, and the limit $j\to\infty$ takes us to the classical regime. The exactly solvable 2- to 4-qubit systems support benchmarking of quantum simulations and experiments, especially in quantum devices where higher-qubit implementations face problems. In general, one has to resort to numerical simulations for higher-qubit systems. Here, parity (or collective $\pi$ rotation) of the Floquet operator plays a crucial role in simplifying the numerical simulations. The Floquet operator satisfies the following property:
\begin{align}\label{Eq:symmetry:sigmay}
    \left[\mathcal{U}, \otimes^{N}_{l=1}\sigma_l^y\right] = 0.
\end{align}
The eigenstates of $\otimes^{N}_{l=1}\sigma_l^y$, with $N=2j$ representing the number qubits, are given as follows:
\begin{align}
    \otimes^{N}_{l=1}\sigma^y_l |\Phi_q^\pm \rangle = \pm |\Phi_q^\pm \rangle \quad &\text{where} \quad |\Phi_q^{\pm}\rangle = \frac{1}{\sqrt{2}} \left(|W_q\rangle \pm i^{N - 2q}|\overline{W}_q\rangle\right), \\
    |W_q\rangle = {\begin{pmatrix} N \\ q \end{pmatrix}}^{-\frac{1}{2}} \sum_{\mathcal{P}} {\left( \otimes^q | 1\rangle \otimes^{N-q}|0\rangle \right)}_{\mathcal{P}} \quad &\text{and} \quad 
    |\overline{W}_q\rangle = {\begin{pmatrix} N \\ q \end{pmatrix}}^{-\frac{1}{2}} \sum_{\mathcal{P}} {\left( \otimes^q | 0\rangle \otimes^{N-q}|1\rangle \right)}_{\mathcal{P}}.
\end{align}
Here, $\mathcal{P}$ denotes the sum over all possible permutations of the qubits, $0 \leq q \leq (N-1)/2$ for odd $N$ and $0 \leq q \leq N/2$ for even $N$, excluding the last state $|\Phi_{j}^-\rangle$. The $(N+1)$-dimensional Hilbert space arises by restricting the full $2^{N}$-dimensional qubit space to permutation-symmetric spin-$j$ sector. Then, parity further block diagonalizes this sector.

The Floquet operator takes a block diagonal form in these bases. In the case of odd-$N$, the Floquet operator $\mathcal{U}$ can be expressed as follows:
\begin{align}
    \mathcal{U} = \begin{pmatrix} 
        \mathcal{U}_+ & 0\\
        0 &\mathcal{U}_-
    \end{pmatrix}.
\end{align}
Whereas, for the case of even-$N$, the state $| \Phi_0^+\rangle$ is an eigenstate with eigenvalue $\langle \Phi_0^+ | \mathcal{U} | \Phi_0^+\rangle = -1$. Therefore, we get the following block diagonal form:
\begin{align}\label{eq:even-N}
    \mathcal{U} = \begin{pmatrix}
        -1 &0 &0 \\
        0 &\mathcal{U}_+ & 0\\
        0 &0 &\mathcal{U}_-
    \end{pmatrix}  \quad \text{for even-}N.
\end{align}
The standard SU(2) spin-coherent states have minimal uncertainty and are localized in phase space, making them ideal for studying the quantum-classical correspondence. Since the system can be viewed as a multi-qubit interacting system, it provides a platform to study quantum correlations and their time evolution. In the semi-classical limit, this would offer a clean comparison between classical trajectories in the phase space and quantum evolution. The $N$-qubit initial spin-coherent state is given by
\begin{align}\label{Eq:generalstate}
    |\theta_0, \phi_0\rangle = \otimes^{N} \left[\cos\left(\frac{\theta_0}{2}\right) | 0\rangle + e^{-i\phi_0}\sin\left(\frac{\theta_0}{2}\right)| 1 \rangle\right],
\end{align}
where $|0\rangle = {\left[1, 0\right]}^T$ and $|1\rangle = {\left[0, 1\right]}^T$. This state can also be written in the $|j, m\rangle$ basis as $ |\theta_0, \phi_0\rangle = \sum_{m = -j}^{j} \langle j, m | \theta_0, \phi_0\rangle |j, m\rangle $~\cite{Bandyopadhyay2004} and coefficients are given by
\begin{align}
    \langle j, m | \theta_0, \phi_0\rangle =  \cos^{2j}\left(\frac{\theta_0}{2}\right) {\left[e^{-i \phi_0} \tan\left(\frac{\theta_0}{2}\right)\right]}^{j-m} \sqrt{\begin{pmatrix}
        2j\\
        j-m
    \end{pmatrix}}.
\end{align}
This transformation allows us to interchange between the qubit basis and the spin basis. Recall that for the total spin $j$, the Floquet operator $\mathcal{U}$ has $N+1$ dimensions. However, the initial spin-coherent state has dimensions of $2^{N}$. Therefore, this transformation, along with symmetry (see Eq.~\eqref{Eq:symmetry:sigmay}), greatly improves computational efficiency.

There are many approaches to the quantum signatures of chaos in the semi-classical regime. These include spacing statistics, entanglement entropy, OTOC, Loschmidt echo, and others. In the deep quantum regime, however, only a few of these approaches, such as entanglement entropy and OTOC, are effective. Since the entanglement entropy allows us to study the generation of quantum correlations right from the deep quantum regime to the semi-classical regime, it is discussed in detail. The spacing statistics is a robust and well-established indicator of quantum chaos in the semi-classical regime. Therefore, the spacing statistics is discussed for higher-qubit systems. Other quantum signatures of chaos are also discussed briefly in the later section.

From the classical perspective, the OTOC is one of the natural approaches. It is given by~\cite{sreeram2021out}
\begin{align}\label{Eq:otoc}
    C_\rho(t) = -\frac{1}{2} \tr \left(\rho {\left[A(t), A(0)\right]}^2 \right), \; \text{where}\; A(t) = \mathcal{U}^{-t}A(0)\mathcal{U} \;\; \text{and}\; \rho = \frac{1}{N+1}\mathds{I}.
\end{align}
The OTOC is an indicator of information scrambling. Here, initially localized operator gets entangled with other one-particle operators on other sites. This leads to the loss of information of the initial state. In the semi-classical limit, the OTOC is expected to grow exponentially, i.e. $\exp\left(2\lambda_Q t\right)$. Here, the quantum Lyapunov exponent $\lambda_Q$ is defined from the exponential growth rate of underlying commutator $[A(t),A(0)]\sim e^{\lambda_Q t}$ in the semiclassical limit~\cite{rozenbaum2017lyapunov, trunin2023refined, majewski2006quantum, falsaperla2002quantum}. This exponential growth lasts until time reaches the Ehrenfest time $t_E \approx \log(N+1)/\lambda$, beyond which it gets saturated or oscillates due to finite-Hilbert space effect. Even if we take large number of qubits, let us say $N \approx 10^{10}$. The Ehrenfest time $t_E$ still remains of the order of $10$. Since this short-term growth can be deceptive, only the above-mentioned approaches are discussed in the chapter.

\subsection{Two-Qubits System}\label{sec:sub:two-qubit}
The 2-qubit realization of the QKT corresponding to spin $j=1$ represents the minimal system with non-trivial quantum correlations and nonlinear dynamics. The basis states for this case are defined as follows:
\begin{align}
    |\Phi_0^\pm\rangle &= \frac{1}{\sqrt{2}} |00\rangle \mp \frac{1}{\sqrt{2}} |11\rangle = {\left[\frac{1}{\sqrt{2}}, 0, 0, \mp\frac{1}{\sqrt{2}}\right]}^T \quad \text{and} \\
    |\Phi_1^+\rangle &= \frac{1}{\sqrt{2}} |10\rangle + \frac{1}{\sqrt{2}} |01\rangle = {\left[0, \frac{1}{\sqrt{2}}, \frac{1}{\sqrt{2}}, 0\right]}^T.
\end{align}
This allows us to write the initial spin-coherent state in the above basis as follows:
\begin{align}
|\psi_0\rangle = \frac{1}{\sqrt{2}}
\Big\{
& e^{-i\phi_0} \left(\cos\theta_0 \cos\phi_0 + i \sin\phi_0\right) |\Phi_0^+\rangle
+ e^{-i\phi_0} \sin\phi_0 \, |\Phi_1^+\rangle \notag
\\
& \quad
+ \left[\cos^2\left(\frac{\theta_0}{2}\right) + i \sin^2\left(\frac{\theta_0}{2}\right)\right] |\Phi_0^-\rangle
\Big\}.
\end{align}
The $n$-th power of the Floquet operator in the $\Phi$-basis is given by
\begin{equation}
    \mathcal{U}^n = \begin{bmatrix}
        \cos\left(\frac{n\pi}{2}\right) & -\sin\left(\frac{n\pi}{2}\right)e^{-ik} & 0 \\
        \sin\left(\frac{n\pi}{2}\right)e^{ik} & \cos\left(\frac{n\pi}{2}\right) & 0 \\
        0 & 0 & e^{-i n k}
    \end{bmatrix}.
\end{equation}
Thus, the time-evolved state $|\psi_n\rangle = \mathcal{U}^n |\psi_0\rangle$ can be expressed as follows:
\begin{equation}
    |\psi_n\rangle = c_0 |\Phi_0^+\rangle + c_1 |\Phi_1^+\rangle + c_2 |\Phi_0^-\rangle,
\end{equation}
where the coefficients $c_0$, $c_1$, and $c_2$ are given by
\begin{align}
    c_0 &= \frac{e^{-i (k + \phi_0)}}{\sqrt{2}}  \big[-\sin\left(\frac{n\pi}{2}\right)\sin(\theta_0) + e^{ik} \cos\left(\frac{n\pi}{2}\right)\big(\cos(\theta_0)\cos(\phi_0) + i \sin(\phi_0)\big)\big], \label{eq:c0} \\
    c_1 &= \frac{e^{-i\phi_0}}{\sqrt{2}}  \big[\cos\left(\frac{n\pi}{2}\right)\sin(\theta_0) + e^{ik} \sin\left(\frac{n\pi}{2}\right)\big(\cos(\theta_0)\cos(\phi_0) + i \sin(\phi_0)\big)\big] \label{eq:c1} \text{ and } \\
    c_2 &= \frac{e^{-i(n k + \phi_0)}}{\sqrt{2}}  \big(\cos(\phi_0) + i \cos(\theta_0)\sin(\phi_0)\big) \label{eq:c2}.
\end{align}
Although we can construct a two-qubit density matrix in the computational basis:
\begin{equation}
    \rho_{12}(n) = \frac{1}{2}\begin{bmatrix}
         |c_0 + c_2|^2 & (c_0 + c_2)c_1^* & (c_0 + c_2)c_1^* & -(c_0 + c_2)(c_0 - c_2)^* \\
         (c_0 + c_2)^*c_1 & |c_1|^2 & |c_1|^2 & -c_1(c_0 - c_2)^* \\
         (c_0 + c_2)^*c_1 & |c_1|^2 & |c_1|^2 & -c_1(c_0 - c_2)^* \\
         -(c_0 - c_2)(c_0 + c_2)^* & -(c_0 - c_2)c_1^* & -(c_0 - c_2)c_1^* & 1 - |c_1|^2 - c_2c_0^* -c_2^* c_0
    \end{bmatrix},
\end{equation}
the single-qubit RDM is of special interest to us. This is because it addresses several questions: (1) it would tell us whether one qubit sees the rest as an effective environment, (2) how fast the information gets scrambled, and (3) how does this scrambling depend on whether the underlying classical dynamics is regular or chaotic. From a practical viewpoint, the single-qubit RDMs are easily accessible in experiments. A single-qubit linear entropy is the standard entanglement probe in QKT realizations with NMR~\cite{krithika2019nmr} or superconducting qubits~\cite{neil2016ergodic}. It measures the entanglement of that qubit with the rest.

Now, we obtain the single-qubit RDM by carrying out a partial trace over the other qubit as follows:
\begin{align}\label{Eq:2qubit:rho}
    \rho_1(n) = \begin{bmatrix}
        \frac{1}{2} + \Re[c_0 c_2^*] & \Re[c_1 c_2^*] + i \Im[c_0 c_1^*] \\
        \Re[c_1 c_2^*] - i \Im[c_0 c_1^*] & \frac{1}{2} - \Re[c_0 c_2^*]
    \end{bmatrix}.
\end{align}
Its eigenvalues are $R_\pm = \frac{1}{2} \pm \sqrt{\Re[c_0 c_2^*]^2 + \Im[c_0 c_1^*]^2 + \Re[c_1 c_2^*]^2}$ and eigenvectors are given by 
\begin{align}
    { \left[ \frac{ \Re[c_0 c_2^*] \pm \sqrt{{ \Re[c_0 c_2^*]}^2  +  {\Re[c_1 c_2^*]}^2 + {\Im[c_0 c_1^*]}^2 }}{\Re[c_1 c_2^*] - i \Im[c_0 c_1^*]} , 1 \right] }^T.
\end{align}
As mentioned earlier, the linear entropy corresponding to the single-qubit RDM is preferred for the analysis. It is given by
\begin{align}
    S_{(\theta_0, \phi_0)}^{(2)}(n, k) = 2R_+ R_- = \frac{1}{2} - 2\big({\Re[c_0 c_2^*]}^2 + {\Im[c_0 c_1^*]}^2 + {\Re[c_1 c_2^*]}^2\big).
\end{align}
The coefficients $\Re[c_0 c_2^*]$, $\Im[c_0 c_1^*]$, and $\Re[c_1 c_2^*]$ can be obtained using Eqs.~\eqref{eq:c0},~\eqref{eq:c1} and~\eqref{eq:c2}. The long-time averaged quantities effectively coarse-grain quantum fluctuations and reveal how classical chaos imprints on quantum dynamics. Computing the infinite-time average linear entropy given by
\begin{align}
    \langle S^{(2)}_{(\theta_0, \phi_0)}(k) \rangle = \lim_{N\to\infty} \frac{1}{N} \sum_{n=0}^{N-1} S^{(2)}_{(\theta_0, \phi_0)}(n, k),
\end{align}
we can quanty the degree of ergodicity. This further allows us to distinguish regular from chaotic regions where it approaches characteristic values~\cite{dogra2019quantum} given by 
\begin{align}\label{eq:S_rmt}
    S_\text{RMT} = \frac{N-1}{2N}.    
\end{align} 
For the special initial state $(\theta_0, \phi_0) = (0,0)$, the infinite-averaged linear entropy is $1/4$. For the general initial spin-coherent state, it is given by
\begin{align}\label{Eq:linearentropy}
    \langle S^{(2)}_{(\theta_0, \phi_0)}(k) \rangle
        &= \frac{1}{1024}\Big[106 + 8\cos(2\theta_0) + 14\cos(4\theta_0) + (2\cos(4\theta_0) - 8\cos(2\theta_0) + 6)\cos(4\phi_0) \Big] \notag \\
        &\quad + \frac{1}{128} {\Big[3+\cos(2\theta_0)+2\cos(2\phi_0)\sin^2\theta_0\Big]}^2 + \frac{1}{16}\sin(k)\sin\theta_0\sin(2\theta_0)\sin(2\phi_0) \notag \\
        &\quad + \frac{1}{1024} \left[32\cos(k)\sin^2\theta_0\big(\cos(2\phi_0) \left[3+\cos(2\theta_0)\right] - 2\sin^2\theta_0\big)\right].
\end{align}
Now, the linear entropies for the special states are given by
\begin{align}
    \langle S^{(2)}_{(0,0)}\left(k \right)\rangle = \frac{1}{4} \quad \text{and} \quad
    \langle S^{(2)}_{(\frac{\pi}{2},-\frac{\pi}{2})}\left(k\right)\rangle = \frac{1}{4}\sin^2\left(\frac{k}{2}\right) \quad \text{for }\; 0 < k < 2\pi.
\end{align}
Note that, at $k=0$, there is discontinuity in the entanglement for the initial state $|\theta_0 = 0, \phi_0 = 0 \rangle$. For $k\neq 0$, we can see that the state $|\theta_0 = 0, \phi_0 = 0\rangle$ saturates to its corresponding RMT value. It shows highly quantum behavior. For the state $|\theta_0 = \pi/2, \phi_0 = -\pi/2\rangle$, however, the entanglement depends on the kick strength $k$. For small $k$, it is close to zero and increases sinusoidally to reach RMT value at $k=\pi$. In spite of being in the deep quantum regime, this low entanglement for small values of $k$ indicates non-thermal behavior. This further suggests non-chaotic behavior of this state for small values of $k$.
\begin{figure}
    \includegraphics[width=1\textwidth]{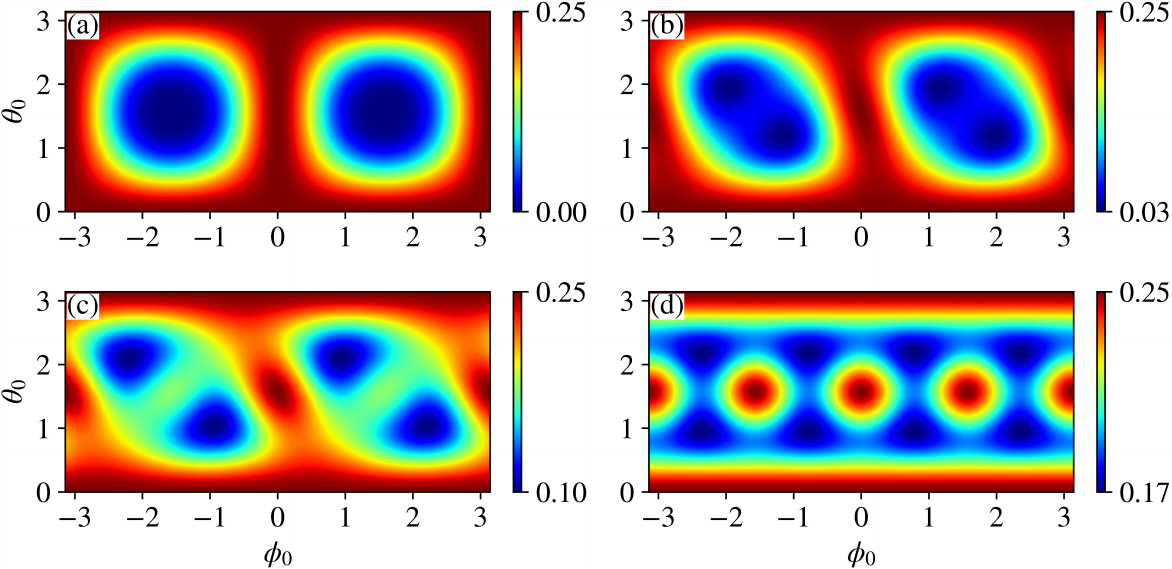}
    \caption{The infinite-time averaged linear entropy plotted for precession angle $p = \pi/2$, $j=1$, taking a grid of $200\times 200$ initial points for each subplot. Panels: (a) $k=0$, (b) $k=1$, (c) $k=2$ and (d) $k=\pi$.}\label{fig:linear_entropy_2}
\end{figure}
The increase of entropy with $k$ indicates that the single-qubit is behaving as if it is kept in a bath, as illustrated in Fig.~\ref{fig:linear_entropy_2}. Interestingly, even in this deep quantum regime, we can observe bifurcation-like feature in the infinite-time averaged entanglement. However, this pronounced bifurcation-like feature seems to appear at $k=2$ itself much earlier than that in the classical phase space (refer Figs.~\ref{fig:pahse_portrait}(a) and (b)).

\subsection{Three-Qubits System}\label{sec:sub:three-qubit}
The 3-qubit realization of the QKT is the minimal system with genuine multipartite entanglement and nonlinear dynamics. The basis states for this system are given as follows: 
\begin{align}
    |\Phi^\pm_0\rangle &= \frac{1}{\sqrt{2}} \left(|000\rangle \pm |111\rangle\right), \;  |\Phi^\pm_1\rangle = \frac{1}{\sqrt{2}} \left(|W\rangle \pm i |\overline{W}\rangle\right),\\
    |W\rangle &= \frac{1}{\sqrt{3}} \sum_{\mathcal{P}} |001\rangle_{\mathcal{P}} \quad \text{and} \quad |\overline{W}\rangle = \frac{1}{\sqrt{3}} \sum_{\mathcal{P}} |110\rangle_{\mathcal{P}},
\end{align}
where $\sum_{\mathcal{P}}$ denotes summation over all possible permutations. The Floquet operator in this basis is given by
\begin{align}
    \begin{split}
 \mathcal{U} &= \begin{pmatrix}
 \mathcal{U}_+    &0\\
            0   &\mathcal{U}_-
        \end{pmatrix} \; \text{where}\;\,
 \mathcal{U}_{\pm} = \pm e^{\mp i\frac{\pi}{4}} e^{-\frac{ik}{6}} \begin{pmatrix}
            \alpha   &\mp \beta^*\\
            \pm \beta   &\alpha^*
        \end{pmatrix}, 
    \end{split}\\
        \alpha &= \frac{i}{2} \exp\left(-i\frac{k}{3}\right) \quad \text{and} \quad \beta = \frac{\sqrt{3}}{2} \exp\left(i\frac{k}{3}\right).
\end{align}
By expressing $\mathcal{U}_+$ as a rotation $e^{-i\gamma \vec{\sigma}\cdot \hat{\eta}}$ by an angle $\gamma$ about an axis $\hat{\eta}$~\cite{dogra2019quantum}, we get
\begin{align}
    e^{-i\gamma \vec{\sigma}\cdot \hat{\eta}} = \begin{pmatrix}
        \cos \gamma - i \sin \gamma \cos \theta &-i \sin \gamma \sin \theta e^{-i\chi} \\
         -i \sin \gamma \sin \theta e^{i\chi}   &\cos \gamma + i \sin \gamma \cos \theta
    \end{pmatrix}.
\end{align}
The expressions for $\gamma$, $\theta$, and $\chi$ are obtained by comparing with the Floquet operator $\mathcal{U}_\pm$ (without global phase factors) as follows:
\begin{align}\label{Eq:gamma3}
    \begin{split}
        \cos\gamma = \frac{1}{2}\sin\left(\frac{k}{3}\right), \; \sin\theta = \frac{\sqrt{3}}{2\sin\gamma}\; \text{and} \; \chi = \frac{k}{3} + \frac{\pi}{2}.
    \end{split}
\end{align}
Then the time-evolved Floquet operator is given by
\begin{align}
 \mathcal{U}^n_{\pm} &= {\left(\pm 1\right)}^n e^{\mp i n \frac{\pi}{4}} e^{-\frac{ik}{6}n} \begin{pmatrix}
        \alpha_n   &\mp \beta^*_n\\
        \pm \beta_n   &\alpha^*_n
    \end{pmatrix} \; \text{where} \\
    \alpha_n &= \cos(n\gamma) + \frac{i}{2} \frac{\sin (n\gamma)}{\sin\gamma}\cos\left(\frac{k}{3}\right) \; \text{and} \;\, \beta_n = \frac{\sqrt{3}}{2} \frac{\sin (n\gamma)}{\sin\gamma} \exp\left(i\frac{k}{3}\right). \label{Eq:3qubit_alphan}
\end{align}
Note that the range of $\cos(\gamma)$ is restricted to $|\cos(\gamma)| \leq 1/2$.

Now, the general initial spin-coherent state is written in the $\Phi$-basis as follows:
\begin{align}
    |\psi_0\rangle =& \frac{1}{\sqrt{2}} \left\lbrace \left[\cos^3\left(\frac{\theta_0}{2}\right) + i e^{-3i\phi_0}\sin^3\left(\frac{\theta_0}{2}\right)\right] |\Phi_0^+\rangle  + \left[\cos^3\left(\frac{\theta_0}{2}\right) - i e^{-3i\phi_0}\sin^3\left(\frac{\theta_0}{2}\right)\right] |\Phi_0^-\rangle \right\rbrace \notag \\
    & + \sqrt{\frac{3}{8}} e^{-2i\phi_0} \sin\left(\frac{\theta_0}{2}\right)\sin(\theta_0) \left\lbrace \left[ e^{i\phi_0}\cot\left(\frac{\theta_0}{2}\right) - i\right] |\Phi_1^+\rangle + \left[e^{i\phi_0}\cot\left(\frac{\theta_0}{2}\right) + i\right] |\Phi_1^-\rangle \right\rbrace.
\end{align}
Using the Floquet operator, the time-evolved state is obtained as follows:
\begin{align}
    |\psi_n\rangle =& c_0' |\Phi_0^+\rangle + c_1' |\Phi_1^+\rangle + c_2' |\Phi_0^-\rangle + c_3' |\Phi_1^-\rangle, 
\end{align}
where the coefficients $c_i'$ are given as follows:
\begin{alignat}{2}
    c_0' =& e^{-\frac{i}{4}(n\pi + 6\phi_0)} c_f \left[2\alpha_n \cos(\theta_0)\cos(\phi_0) - \sqrt{3}\beta_n^* \sin(\theta_0) + i \alpha_n \left(\sin(\theta_0) + 2\sin(\phi_0) \right)\right],  \notag \\
    c_1' =& e^{-\frac{i}{4}(n\pi + 6\phi_0)} c_f \left[2\beta_n \cos(\theta_0)\cos(\phi_0) + \sqrt{3}\alpha_n^* \sin(\theta_0) + i \alpha_n \left(\sin(\theta_0) + 2\sin(\phi_0) \right)\right], \notag \\
    c_2' =& e^{\frac{i}{4}(5n\pi - 6\phi_0)} c_f^* \left[2\alpha_n \cos(\theta_0)\cos(\phi_0) + \sqrt{3}\beta_n^* \sin(\theta_0) - i \alpha_n \left(\sin(\theta_0) - 2\sin(\phi_0) \right)\right], \notag \\
    c_3' =& -e^{\frac{i}{4}(5n\pi - 6\phi_0)} c_f^* \left[2\beta_n \cos(\theta_0)\cos(\phi_0) - \sqrt{3}\alpha_n^* \sin(\theta_0) - i \beta_n \left(\sin(\theta_0) - 2\sin(\phi_0) \right)\right], \notag \\
    \text{and}\;  c_f =& \frac{1}{2\sqrt{2}}\left[\cos\left(\frac{\theta_0 + \phi_0}{2}\right)-i\sin\left(\frac{\theta_0 - \phi_0}{2}\right)\right].
\end{alignat}
Then, the single qubit RDM $\rho_1(n) = \tr_{23} |\psi_n\rangle \langle \psi_n|$ is given by
\begin{align}\label{eq:3qubitrho}
    \rho_1(n) =& \begin{pmatrix}
        \frac{1}{2} + \Re[c_0' {c_2'}^*] + \frac{1}{3}\Re[c_1' {c_3'}^*]   &a_2 + a_4 \\
        {(a_2 + a_4)}^*   &\frac{1}{2} - \Re[c_0' {c_2'}^*] - \frac{1}{3}\Re[c_1' {c_3'}^*]
    \end{pmatrix} \; \text{where} \\
    a_2 + a_4 =& -\frac{i}{3} (c_1' + c_3'){(c_1' - c_3')}^* + \frac{\sqrt{3}}{6} (c_0' + c_2') {(c_1' + c_3')}^* -\frac{1}{2\sqrt{3}} (c_1' - c_3'){(c_0' - c_2')}^* \notag.
\end{align}
This further allows us to write the linear entropy as follows:
\begin{align}\label{eq:S3qubit}
    S^{(3)}_{(\theta_0, \phi_0)}(n, k) = \frac{1}{2} - 2 {\left(\Re[c_0' {c_2'}^*] + \frac{1}{3}\Re[c_1' {c_3'}^*]\right)}^2 - 2 {|a_2 + a_4|}^2.
\end{align}
\begin{figure}
    \includegraphics[width=1\textwidth]{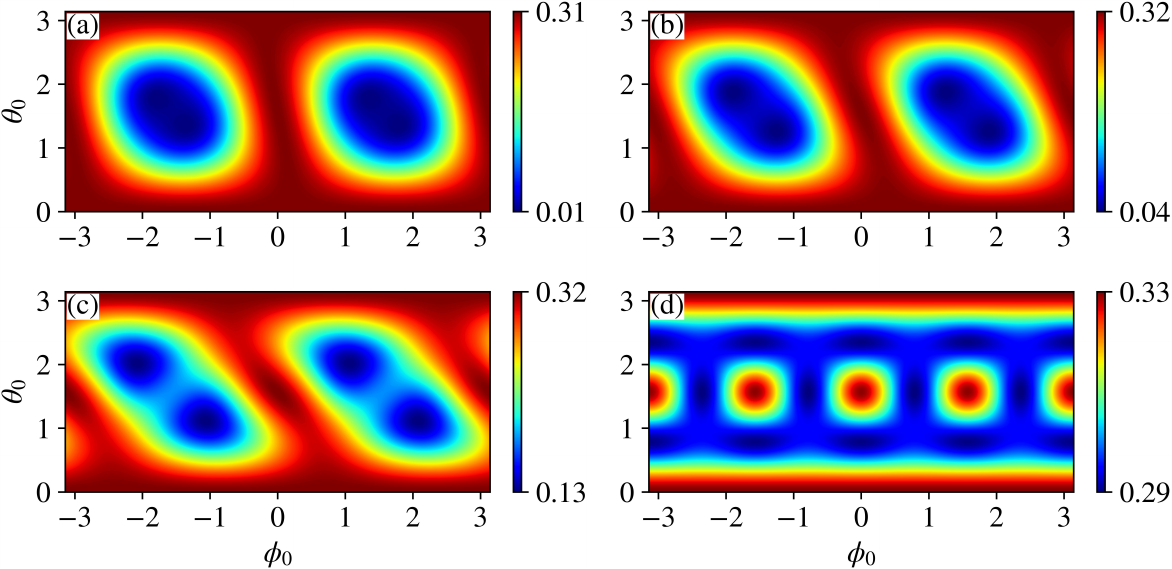}
    \caption{The long-time averaged linear entropy plotted for precession angle $p = \pi/2$, $j=3/2$, taking a grid of $200\times 200$ initial points for each subplot. Panels: (a) $k=0.5$, (b) $k=1$, (c) $k=2$ and (d) $k=3\pi/2$.}\label{fig:linear_entropy_3}
\end{figure}
The analytical calculation of the infinite-time averaged linear entropy for the general initial spin-coherent state is cumbersome. Since the analysis for the special states (mentioned in the two-qubits case) serves the purpose, only those initial states are considered here. The infinite-time averaged linear entropy for these states are given by
\begin{align}
    \langle S^{(3)}_{(0,0)}(k) \rangle =  \frac{5 - 2\sin^2\left(\dfrac{k}{3}\right)}{{\left[4 - \sin^2\left(\dfrac{k}{3}\right)\right]}^2} \quad \text{for } \; 0 < k < 3\pi,
\end{align} 
and
\begin{align}
    \langle S^{(3)}_{(\frac{\pi}{2}, -\frac{\pi}{2})}(k) \rangle = \frac{8 - 5\sin^2\left(\dfrac{k}{3}\right)}{{\left[4 - \sin^2\left(\dfrac{k}{3}\right)\right]}^2} \sin^2\left(\frac{k}{3}\right).
\end{align}
The state $|\theta_0 = 0, \phi_0 = 0\rangle$ tends to have higher entanglement, similar to the case of a 2-qubit system. The state $|\theta_0 = \pi/2, \phi_0 = -\pi/2\rangle$, on the other hand, has low entanglement for small $k$ which goes on increasing till $k=3\pi/2$, where it approaches Eq.\eqref{eq:S_rmt}. To get the full picture, the long-time averaged linear entropy is illustrated in Fig.~\ref{fig:linear_entropy_3}. The bifurcation-like process develops more slowly (as $k$ is increased) than in the 2-qubit system. This consistency provides insight into how quantum effects shape the nonlinear dynamics.

\subsection{Four-Qubits System}\label{sec:sub:four-qubit}
This system is particularly interesting as it is exactly solvable and exhibits signatures of ergodicity and thermalization~\cite{dogra2019quantum}. The basis states for this case are given by
\begin{align}
    |\Phi_0^\pm\rangle =& \frac{1}{\sqrt{2}} |0000\rangle \pm \frac{1}{\sqrt{2}} |1111\rangle = \frac{1}{\sqrt{2}}{\left[ 1, 0,  0,  0,  0,  0,  0,  0,  0,  0,  0,  0,  0,  0,  0,  \pm 1\right]}^T, \\
    |\Phi_1^\pm\rangle =& \frac{1}{\sqrt{2}} |W\rangle \mp \frac{1}{\sqrt{2}} |\overline{W}\rangle = \frac{1}{2 \sqrt{2}}{\left[0, 1, 1, 0, 1, 0, 0, \mp 1, 1, 0, 0, \mp 1, 0, \mp 1, \mp 1, 0\right]}^T, \\ 
    |\Phi_2^+\rangle =& \frac{1}{\sqrt{6}} \sum_{\mathcal{P}}  |0011\rangle = \frac{1}{\sqrt{6}} {\left[0, 0, 0, 1, 0, 1, 1, 0, 0, 1, 1, 0, 1, 0, 0, 0\right]}^T.
\end{align}
The general spin-coherent state is rewritten in these basis states as follows:
\begin{align}
    |\psi_0\rangle =& \frac{1}{\sqrt{2}} \left[\cos^4\left(\frac{\theta_0}{2}\right) + e^{-4i\phi_0}\sin^4\left(\frac{\theta_0}{2}\right)\right] |\Phi_0^+\rangle + \frac{1}{\sqrt{2}} \left[\cos^4\left(\frac{\theta_0}{2}\right) - e^{-4i\phi_0}\sin^4\left(\frac{\theta_0}{2}\right)\right] |\Phi_0^-\rangle \notag \\
    & + \frac{1}{\sqrt{2}} e^{-2i\phi_0} \sin(\theta_0) \left[\cos(\theta_0) \cos(\phi_0)+ i \sin(\phi_0)\right] |\Phi_1^+\rangle + \sqrt{\frac{3}{8}} e^{-2i\phi_0} \sin^2(\theta_0) | \Phi_2^+ \rangle  \notag \\
    & + \frac{1}{\sqrt{2}} e^{-2i\phi_0} \sin(\theta_0) \left[ \cos(\phi_0) + i \cos(\theta_0) \sin(\phi_0)\right] |\Phi_1^-\rangle.
\end{align}
The Floquet operator in these basis states can also be expressed using a procedure similar to the previous cases as follows:
\begin{align}
    \mathcal{U}_+ = -i e^{-\frac{i k}{4}} \begin{pmatrix}
        \alpha' &i{\beta'}^*\\
        i\beta' &{\alpha'}^*
    \end{pmatrix}  \; \text{and} \;\,
    \mathcal{U}_- = \begin{pmatrix}
        0   &1\\
        -e^{-\frac{3}{4}i k}    &0
    \end{pmatrix}.
\end{align}
Here, $\alpha'$ and $\beta'$ are given by
\begin{align}
    \alpha' = \frac{i}{2} e^{-i\frac{k}{2}} \; \text{and} \;
    \beta' = \frac{\sqrt{3}}{2} e^{i\frac{k}{2}}.
\end{align}
Recall that the actual $\mathcal{U}_{+}$ has extra one row and one column than the corresponding $\mathcal{U}_{-}$ with eigenvalue $-1$ in the state $|\Phi_{0}^{+}$ for even-$N$. Because of its trivial action, we will consider only rest of the part to get $\mathcal{U}^{n}$. Further, the time-evolved Floquet operator is given by
\begin{align}
    \mathcal{U}_+^n = e^{-\frac{i}{4}n (k + 2\pi)} \begin{pmatrix}
        \alpha'_n &i{\beta'_n}^*\\
        i\beta'_n &{\alpha'_n}^*
    \end{pmatrix} \; \text{and} \;\,
    \mathcal{U}_-^n = e^{-\frac{3i}{8} n k} \begin{pmatrix}
        \cos\left(\frac{n\pi}{2}\right) &\sin\left(\frac{n\pi}{2}\right)e^{\frac{3i}{8} k}\\
        -\sin\left(\frac{n\pi}{2}\right) e^{-\frac{3i}{8} k} &\cos\left(\frac{n\pi}{2}\right)
    \end{pmatrix},
\end{align}
where $\alpha'_n$ and $\beta'_n$ are given by
\begin{align}\label{Eq:4qubit_alphan}
    \alpha'_n = \cos(n \gamma) + \frac{i}{2} \frac{\sin(n\gamma)}{\sin(\gamma)} \cos\left(\frac{k}{2}\right) \;
    \beta'_n = \frac{\sqrt{3}}{2} \frac{\sin(n\gamma)}{\sin(\gamma)} e^{i\frac{k}{2}} \; \text{and}\; \cos\gamma = \frac{1}{2}\sin\left(\frac{k}{2}\right).
\end{align}
Then the time-evolved states can be given by
\begin{align}
    |\psi_n\rangle =& c_0'' |\Phi_0^+\rangle + c_1'' |\Phi_1^+\rangle + c_2'' |\Phi_2^+\rangle + c_3'' |\Phi_0^-\rangle + c_4'' |\Phi_1^-\rangle,
\end{align}
where the coefficients $c_i^{\prime\prime}$ are given by
\begin{align}
    c_0'' =& \; \frac{e^{-\frac{i}{2}(n\frac{k}{2}+n\pi+8\phi_0)}}{2\sqrt{2}}  \left\{2\alpha_n \left[\cos^4\left(\frac{\theta_0}{2}\right) + e^{-4i\phi_0}\sin^4\left(\frac{\theta_0}{2}\right)\right] + i \sqrt{3} e^{2i\phi_0} \beta_n^* \sin^2(\theta_0) \right\}, \\
    c_1'' =& \; {(-1)}^n \frac{e^{-2i\phi_0}}{\sqrt{2}}  \sin(\theta_0) \left[\cos(\theta_0) \cos(\phi_0)+ i \sin(\phi_0)\right], \\
    c_2'' =& \; \frac{e^{-\frac{i}{2}(n\frac{k}{2}+n\pi+8\phi_0)}}{2\sqrt{2}}  \left\{2i\beta_n \left[e^{4i\phi_0}\cos^4\left(\frac{\theta_0}{2}\right) + \sin^4\left(\frac{\theta_0}{2}\right)\right] + \sqrt{3} e^{2i\phi_0} \beta_n^* \sin^2(\theta_0) \right\}, \\
    c_3'' =& \; \frac{e^{-\frac{i}{4}\left(\frac{3k}{2}(n+1) + 16\phi_0\right)}}{\sqrt{2}} \left\{ e^{\frac{3ik}{8}} \cos\left(\frac{n\pi}{2}\right) \left[e^{4i\phi_0}\cos^4\left(\frac{\theta_0}{2}\right) - \sin^4\left(\frac{\theta_0}{2}\right)\right]  \right. \notag \\ 
    &\quad \left.- \; e^{2i\phi_0} \sin\left(\frac{n\pi}{2}\right) \sin(\theta_0) \left[\cos(\phi_0) + i \cos(\theta_0)\sin(\phi_0)\right] \right\} \quad \text{and} \quad \\
    c_4'' =& \; \frac{e^{-\frac{3ink}{8}}}{\sqrt{2}}  \left\{ e^{\frac{3ik}{8}} \cos\left(\frac{n\pi}{2}\right) \left[\cos^4\left(\frac{\theta_0}{2}\right) - e^{-4i\phi_0}\sin^4\left(\frac{\theta_0}{2}\right)\right] \right. \notag \\ 
    &\quad \left. + \; e^{-2i\phi_0} \cos\left(\frac{n\pi}{2}\right) \sin(\theta_0) \left[\cos(\phi_0) + i \cos(\theta_0)\sin(\phi_0)\right] \right\}.
\end{align}
The single-qubit RDM is obtained by taking partial trace over any three qubits as follows:
\begin{align}
    \rho_1(n) = \begin{pmatrix}
        \frac{1}{2}\left(1+2\,\Re[c_0'' {c_3''}^*]+2\,\Re[c_1'' {c_4''}^*]\right) &p_{12}'' \\
        {p_{12}''}^*   &\frac{1}{2}\left(1-2\,\Re[c_0'' {c_3''}^*]-2\,\Re[c_1'' {c_4''}^*]\right)
    \end{pmatrix},
\end{align}
where,
\begin{align}
    p_{12}'' = \frac{1}{4}\left[(c_1''-c_4'') {(c_3''-c_0'')}^* + (c_0''+c_3'') {(c_1''+c_4'')}^*\right] + \frac{\sqrt{3}}{4}\left[c_2'' {(c_4''-c_1'')}^* + {c_2''}^* (c_4''+c_1'')\right]. \notag
\end{align}
Then the corresponding linear entropy is given by
\begin{align}\label{eq:S4qubit}
    S^{(4)}_{(\theta_0,\phi_0)}(n, k) =& \frac{1}{2} - 2 {\left(\Re[c_0'' {c_3''}^*] + \Re[c_1'' {c_4''}^*]\right)}^2  - 2 {|p_{12}''|}^2.
\end{align}
\begin{figure}
    \includegraphics[width=1\textwidth]{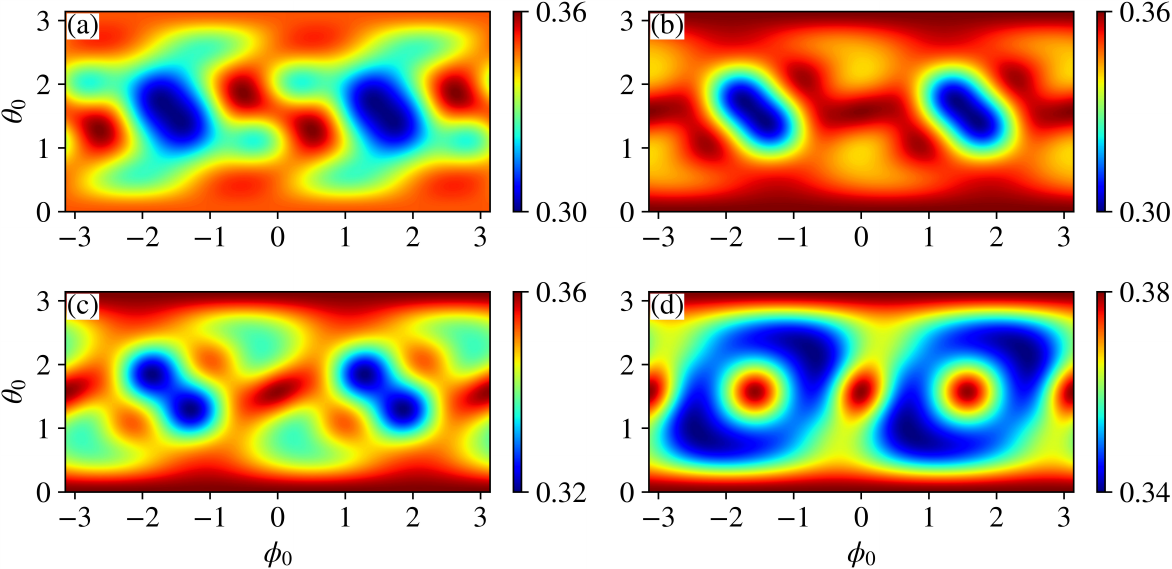}
    \caption{The long-time averaged linear entropy plotted for precession angle $p = \pi/2$, $j=2$, taking a grid of $200\times 200$ initial points for each subplot. Panels: (a) $k=0.7$, (b) $k=1.5$, (c) $k=2$ and (d) $k=2\pi$.}\label{fig:linear_entropy_4}
\end{figure}
The infinite-time averaged linear entropies for the special states are given by
\begin{align}
    \langle S^{(4)}_{(0,0)}(k) \rangle =  \frac{9 + 2\cos^2\left(\dfrac{k}{2}\right)}{8\left[3 + \cos^2\left(\dfrac{k}{2}\right)\right]} \; \text{and}\;
    \langle S^{(4)}_{(\frac{\pi}{2}, -\frac{\pi}{2})}(k) \rangle = \frac{9 - \cos^2\left(\dfrac{k}{2}\right)}{8\left[3 + \cos^2\left(\dfrac{k}{2}\right)\right]}\;\,
    \text{for} \; 0 < k < 4\pi.
\end{align}
Both states show discontinuity at $k=0$. The entanglement is lower for small values of $k$ and approaches value given by Eq.~\eqref{eq:S_rmt} at $k=\pi$. With increasing system size, many interesting structures arise in entanglement landscapes for different values of kicking strength that do not necessarily have a corresponding classical counterpart. For the state $|\theta_0 = \pi/2, \phi_0 = \pm \pi/2 \rangle$, a gradual development of bifurcation is observed in the entanglement landscape (Fig.~\ref{fig:linear_entropy_4}). In the 2-qubit system, the low-entanglement regions are fully separated, whereas in the 4-qubit system they remain connected at $k=2$. For states corresponding to chaotic trajectories, the long-time averaged entanglement approaches the random-state value given by Eq.~\eqref{eq:S_rmt}, consistent with the random permutation-symmetric ensemble~\cite{dogra2019quantum}. As discussed in the following section, in the semiclassical regime the initial growth of entanglement is connected to the instability of the underlying classical dynamics, with the entanglement growth rate governed by the local Lyapunov spectrum~\cite{lerose2020bridging}.

The states corresponding to the period-4 fixed points still exhibit high entanglement, as illustrated in Fig.~\ref{fig:linear_entropy_4}(c). This shows that period-$n$ fixed points do not necessarily correspond to low-entanglement states. The entanglement generated in the QKT depends on specific features of the underlying classical dynamics rather than solely on whether the corresponding classical regime is regular or chaotic~\cite{lombardi2011entanglement}. Consequently, a regular regime can, in certain cases, generate entanglement comparable to that found in chaotic regimes.

\subsection{Many-Qubit System}\label{sec:sub:high-spin}
\begin{figure}
    \includegraphics[width=1\textwidth]{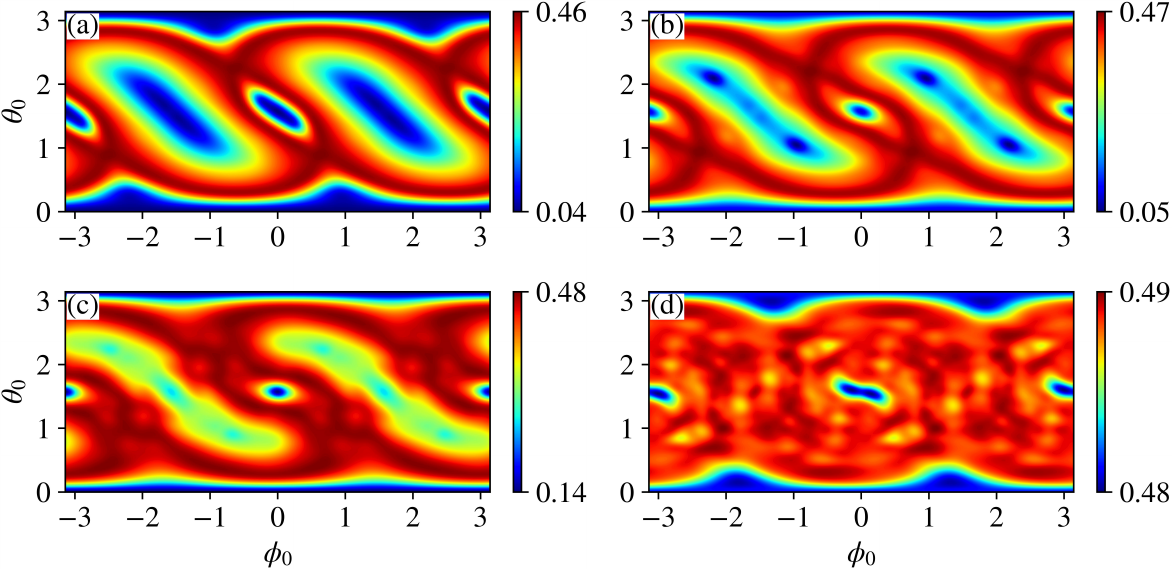}
    \caption{The long-time averaged linear entropy plotted for precession angle $p = \pi/2$, $j=50.5$, taking a grid of $200\times 200$ initial points for each subplot. Panels: (a) $k=2$, (b) $k=2.5$, (c) $k=3$ and (d) $k=6$.}\label{fig:linear_entropy_highspin}
\end{figure}
The earlier analysis of few-qubit systems has shown us the coarse-grained phase space structures of the corresponding trivial fixed points, in particular. To observe fine-grained structures, we need to analyze the system with many qubits. This will also help us understand the interplay between quantum effects and underlying classical chaotic dynamics. With this motivation, the long-time averaged linear entropy for the system of 101 qubits is computed. The linear entropy landscape illustrates fine-grained phase space structures as shown in Fig.~\ref{fig:linear_entropy_highspin}. The blue, extremely low entanglement regions correspond to the trivial fixed points at $k=2$. Unlike the states in a few-qubit systems, the states corresponding to the period-4 fixed points show low entanglement (see Fig.~\ref{fig:linear_entropy_highspin}(a)). It shows us how classical dynamics starts taking shape as the system size is increased. In classical dynamics, these period-4 orbits have chaotic boundaries at $k=2$. The states associated with these boundaries can be seen as red high entanglement regions surrounding the blue low entanglement regions. At $k=2.5$, blue low entanglement regions get bifurcated with green boundaries (see Fig.~\ref{fig:linear_entropy_highspin}(b)). This green boundary region is an indication of a partially stable homoclinic orbit. As $k$ increases, the entanglement increases. At $k=3$, the entanglement of the states corresponding to the period-4 orbits remains lowest compared to other states (see Fig.~\ref{fig:linear_entropy_highspin}(c)). The states corresponding to bifurcated fixed points become green, indicating still lower entanglement compared to the surrounding chaotic states. They have a stable classical counterpart. Interestingly, the states $|\theta_0 = \pi/2, \phi_0 = \pm \pi/2\rangle$ show lower entanglement, potentially even lower than the states corresponding to the bifurcated fixed points. In classical dynamics, the points corresponding to these states are fully chaotic (see Fig.~\ref{fig:LLE}(c)). These points originate from the partially stable homoclinic points present at $k=2$. Although their classical stability is lost upon increasing $k$ beyond two, the corresponding quantum states retain anomalously low entanglement over a finite range of $k$ (see Figs.~\ref{fig:linear_entropy_highspin}(b) and~\ref{fig:linear_entropy_highspin}(c)). Since regular dynamics is typically associated with slower entanglement generation, whereas chaotic dynamics promotes more rapid entanglement growth, this anomalously low entanglement suggests that the quantum dynamics retains a memory of the earlier partially stable homoclinic structure~\cite{Wang2004entanglement, Bandyopadhyay2004entanglement, Bianchi2018}. Therefore, this behavior may be a quantum remnant, or scar-like signature, of the partially stable homoclinic structure. This persistence of low entanglement is intriguing because it survives even after the corresponding classical points become fully chaotic.

\begin{figure}
    \begin{center}
        \includegraphics[width=0.5\textwidth]{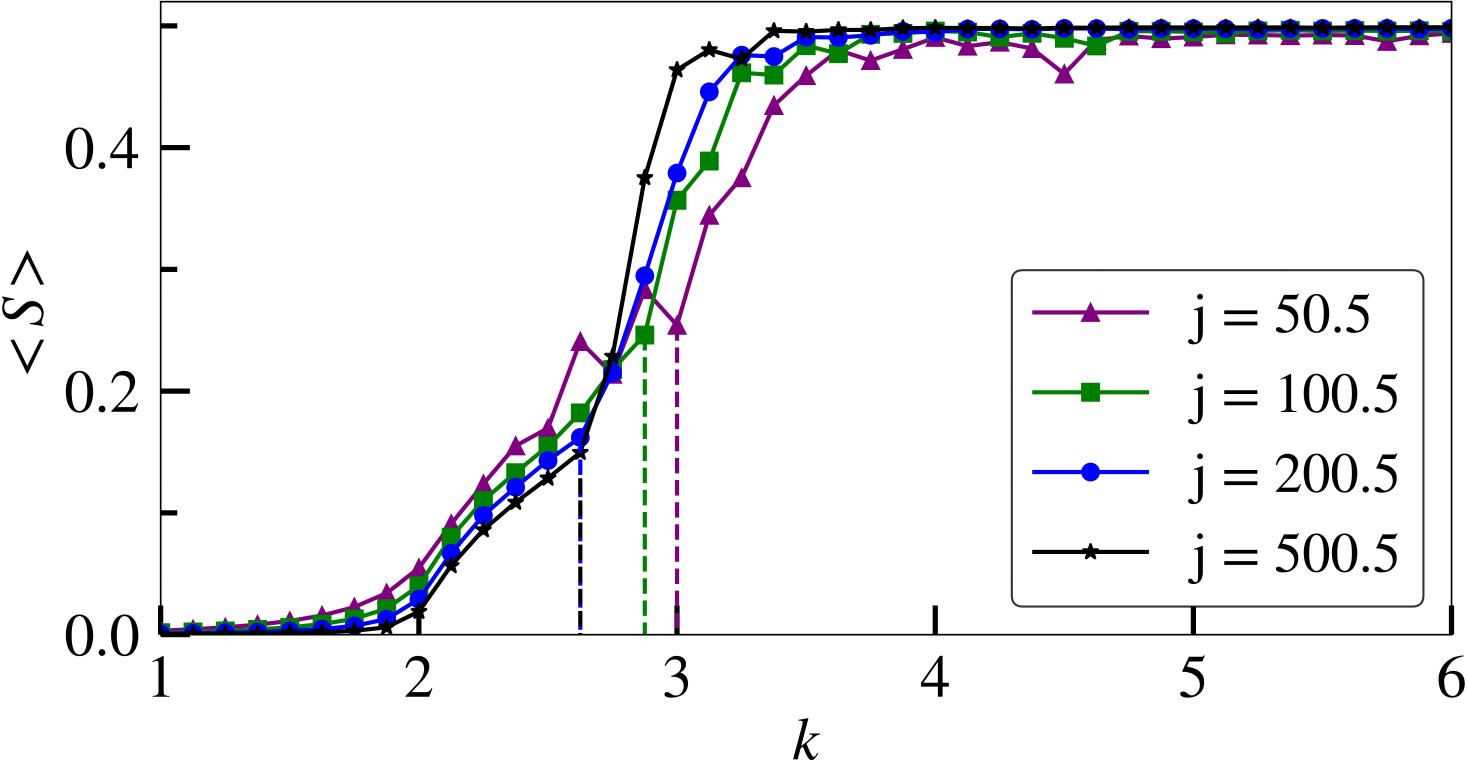}
        \caption{The long-time averaged linear entropy plotted for initial state $|\theta_0 = \pi/2,\phi_0 = -\pi/2\rangle$, precession angle $p = \pi/2$ and $n=1000$.}\label{fig:linear_entropy_vs_k}
    \end{center}
\end{figure}
To further examine the low entanglement of the states $|\theta_0 = \pi/2, \phi_0 = \pm \pi/2\rangle$ at $k=3$, the long-time averaged linear entropy is analyzed as a function of $k$ (see Fig.~\ref{fig:linear_entropy_vs_k}). These states are of special interest as they correspond to homoclinic points formed at $k=2$. In the classical regime, these homoclinic points lose their partial stability almost immediately with a slight increment from $k=2$. In the quantum regime, the entanglement exhibits three distinct domains. The first domain has very low entanglement, the second shows an approximately linear growth, and the third approaches the RMT limit given by Eq.~\eqref{eq:S_rmt}, consistent with the earlier observations~\cite{Wang2004entanglement, Bandyopadhyay2004entanglement, Bianchi2018}. The crossover near $k=2$ is relatively smooth, whereas the crossover from the second to the third domain is sharper. The location of the second crossover shifts toward lower values of $k$ as the system size increases (see Fig.~(23) and related discussion in Ref.~\cite{purohit2025double}). In the classical limit, we expect the two crossovers to merge into a single transition. In the deep quantum regime, however, the state $|\theta_0 = \pi/2, \phi_0 = -\pi/2\rangle$ retains low entanglement over a wider range of $k$. The persistence of this anomalously low entanglement beyond the classical stability regime is therefore consistent with the persistence of signatures of the partially stable homoclinic structure in the quantum dynamics.

As the kicking strength is increased further, the entanglement generally increases and approaches the RMT limit given by Eq.~\eqref{eq:S_rmt} in the fully chaotic regime. However, entanglement is not determined solely by whether the underlying classical dynamics is regular or chaotic. It has been shown that states associated with regular dynamics can exhibit higher entanglement than those associated with chaotic dynamics~\cite{lombardi2011entanglement}. Therefore, entanglement values should be compared with care when considering different initial states and different values of $k$. For example, the states associated with the chaotic boundaries of the period-4 orbits at $k=2$ have entanglement of approximately $0.46$, which is lower than the value of approximately $0.48$ for the states associated with the period-4 orbits at $k=6$ (see Figs.~\ref{fig:LLE}~(a),~\ref{fig:LLE}~(d),~\ref{fig:linear_entropy_highspin}~(a), and~\ref{fig:linear_entropy_highspin}~(d)). Thus, entanglement provides a useful quantum signature of chaotic dynamics, while its quantitative value depends on the specific underlying dynamical structure and does not uniquely distinguish chaotic from regular dynamics.

\subsubsection{Spacing statistics}\label{sec:sub:sub:spacing}
Wigner, in the 1950s, developed RMT to model the spectra of heavy nuclei, as the Hamiltonians of these systems are too complicated and practically impossible to construct. Therefore, instead of trying to compute exact energy levels, he treated the system as a generic random matrix. On studying statistical properties, he showed that we can still extract universal statistical features of their spectra~\cite{wigner1967random}. This success in nuclear physics motivated the extension of RMT to quantum chaos, where complex quantum systems with chaotic classical counterparts exhibit universal spectral statistics matching the RMT predictions. Later,  Bohigas-Giannoni-Schmit (BGS) conjecture, proposed in 1984, extended Wigner's approach to many-body chaotic systems~\cite{bohigas1991random}.
\begin{figure}
    \includegraphics[width=\linewidth]{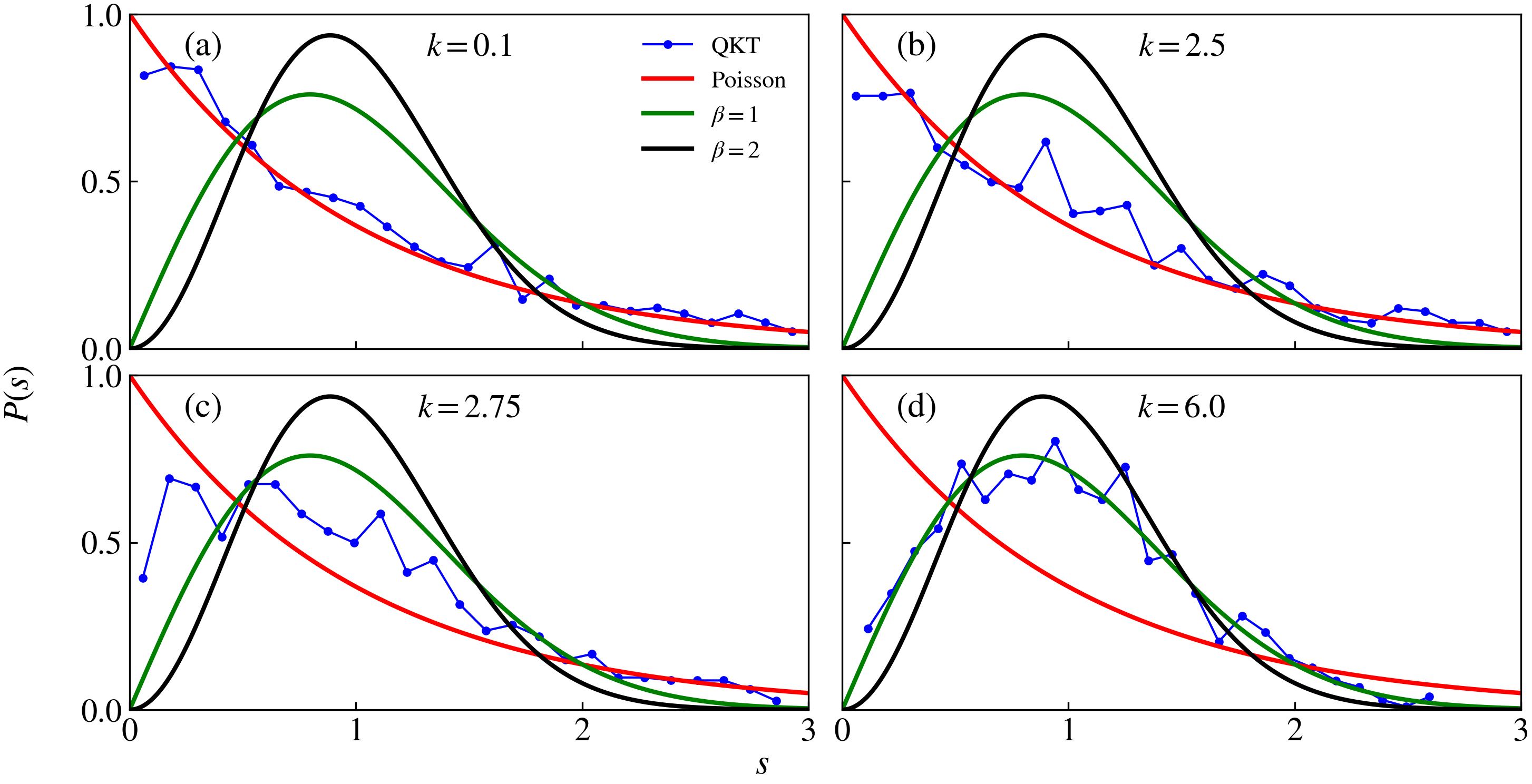}
    \caption{The probability distribution \(P(s)\) versus \(s\) plotted for the QKT with \(p = 1.7 \) and \(j = 1000.5\). Panels: (a) $k=0.1$, (b) $k=2.5$, (c) $k=2.75$ and (d) $k=6.0$.}\label{fig:GOE}
\end{figure}

\begin{figure}
    \includegraphics[width=\linewidth]{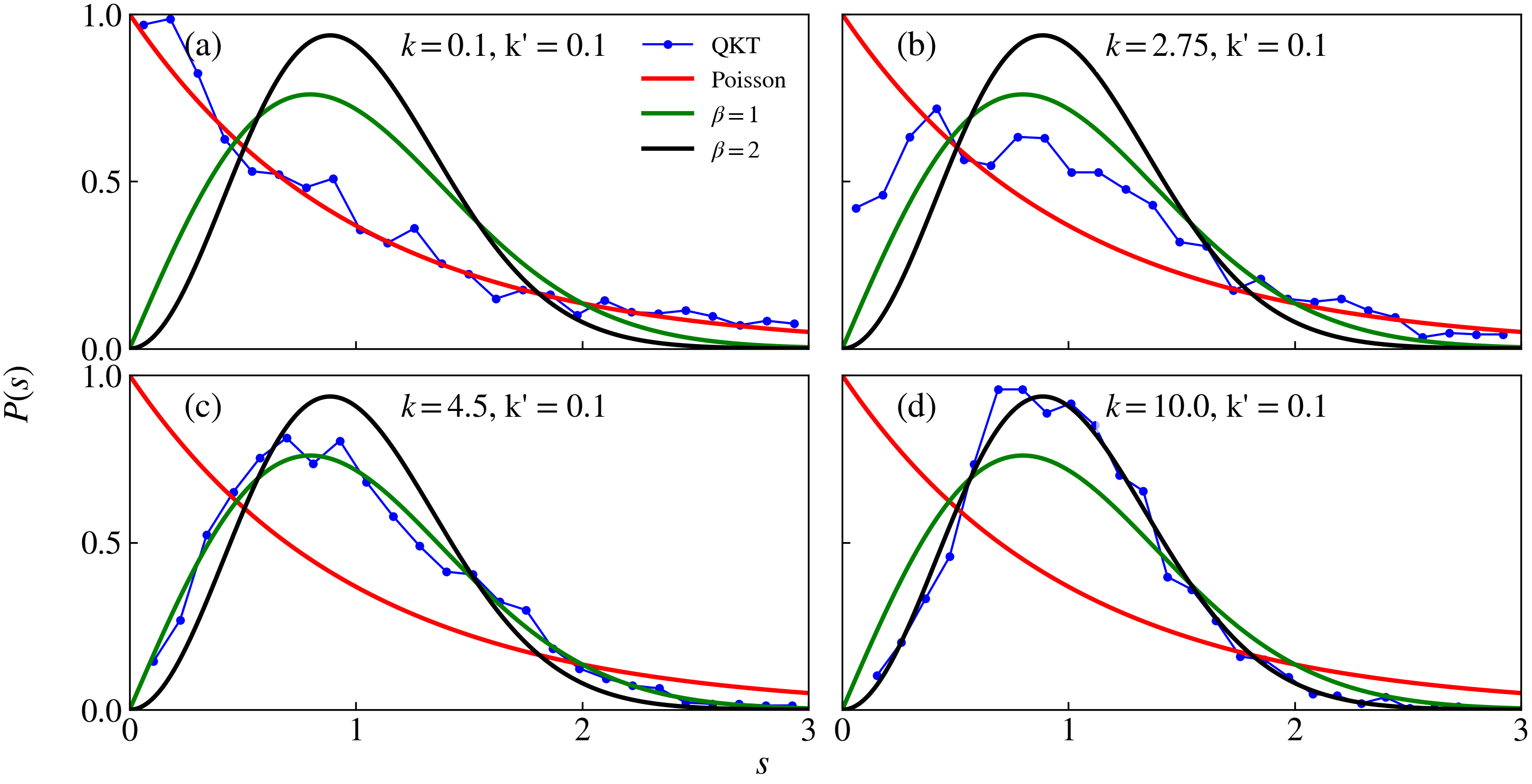}
    \caption{The probability distribution \(P(s)\) versus \(s\) plotted for the broken time-reversal-symmetric extension of the QKT given by Eq.~\eqref{eq:dkt} with \(p = 1.7\), $k' = 0.1$ and \(j = 2000.5\). Panels: (a) $k=0.1$, (b) $k=2.75$, (c) $k=4.5$ and (d) $k=10.0$.}\label{fig:GUE}
\end{figure}

The spacing statistics~\cite{mehta2004random} is rigorous but requires large Hilbert spaces. Therefore, it is applied for high-spin systems and the results are illustrated in Fig.~\ref{fig:GOE}. Finding the probability distribution involves the following six steps:
\begin{enumerate}
    \item Diagonalize the Floquet operator $\mathcal{U}$: $\left\lbrace e^{-iE_l} \right\rbrace$. 
    \item If the system possesses discrete symmetries, separate the quasi-energies into the corresponding symmetry sectors and calculate the spacing statistics within each sector separately.
    \item Arrange quasi-energies $\left\lbrace E_i \right\rbrace$ in ascending order. 
    \item Get spacings: $s_i = E_{i+1}-E_i$.
    \item Unfold: rescale spacings so that the mean spacing is one everywhere.
    \item Plot the histogram of these rescaled quasi-energy spacings.
\end{enumerate}
In the near-integrable regime, $k \leq 2$, the classical phase space is predominantly regular and the quasi-energy levels exhibit approximately Poissonian statistics (see Fig.~\ref{fig:GOE}(a)):
\begin{align}
    P_\text{Poisson}(s) = e^{-s} \quad \text{where}\quad \langle s \rangle = 1.
\end{align}
In the mixed regime, $2 < k < 6$, the spacing statistics evolve between the Poisson and GOE limits, as illustrated in Figs.~\ref{fig:GOE}(b-c). In this regime, the spacing distribution depends on the relative contributions of regular and chaotic regions in the classical phase space and therefore does not correspond to a single universal RMT ensemble. For the time-reversal-symmetric QKT, the fully chaotic regime approaches the Gaussian Orthogonal Ensemble ($\beta=1$) after the relevant symmetries are resolved~\cite{haake1987classical, kus1987symmetry} (see Fig.~\ref{fig:GOE}(d)):
\begin{align}
    P_\text{GOE}(s) = \frac{\pi}{2}s \; e^{-\frac{\pi}{4} s^2} \quad \text{where}\quad \langle s \rangle = 1.
\end{align}
Breaking time-reversal symmetry in the QKT changes the relevant RMT class from GOE to Gaussian Unitary Ensemble ($\beta=2$) in the fully chaotic regime. Accordingly, the spacing distribution approaches the GUE prediction:
\begin{align}
    P_\text{GUE}(s) = \frac{32}{\pi^2}s^2 \; e^{-\frac{4}{\pi} s^2} \quad \text{where}\quad \langle s \rangle = 1.
\end{align}
For the broken time-reversal symmetric extension, even a small nonzero value such as $k'=0.1$ leads to GUE, provided the relevant discrete symmetries are properly resolved~\cite{haake1987classical,purohit2025double}. Here, the precession angle is chosen as $p\neq\pi/2$ (see Figs.~\ref{fig:GOE},~\ref{fig:GUE} and relevant discussion in Ref.~\cite{haake1987classical}). This choice avoids additional symmetry-related degeneracies that can obscure the expected level statistics. 

This shows both the usefulness and the limitations of spacing statistics as a diagnostic of quantum chaos. In fully chaotic regimes, after resolving the relevant symmetries, the spectral statistics approach the appropriate RMT ensemble and provide a powerful quantum signature of chaos. In mixed or intermediate regimes, however, the statistics are generally nonuniversal and cannot by themselves uniquely identify the underlying dynamical regime. The case $k'=0.1$ and $k=4.5$ illustrates this point (see Fig.~\ref{fig:GUE}(c)). Although the system lacks time-reversal symmetry and the underlying classical phase space remains mixed, the distribution appears close to the GOE prediction. Thus, agreement with a particular RMT distribution should be interpreted together with the symmetry class and the underlying classical phase space structure.

\subsubsection{Eigenvector-components statistics}\label{sec:sub:sub:eigenvector}
\begin{figure}
    \includegraphics[width=\linewidth]{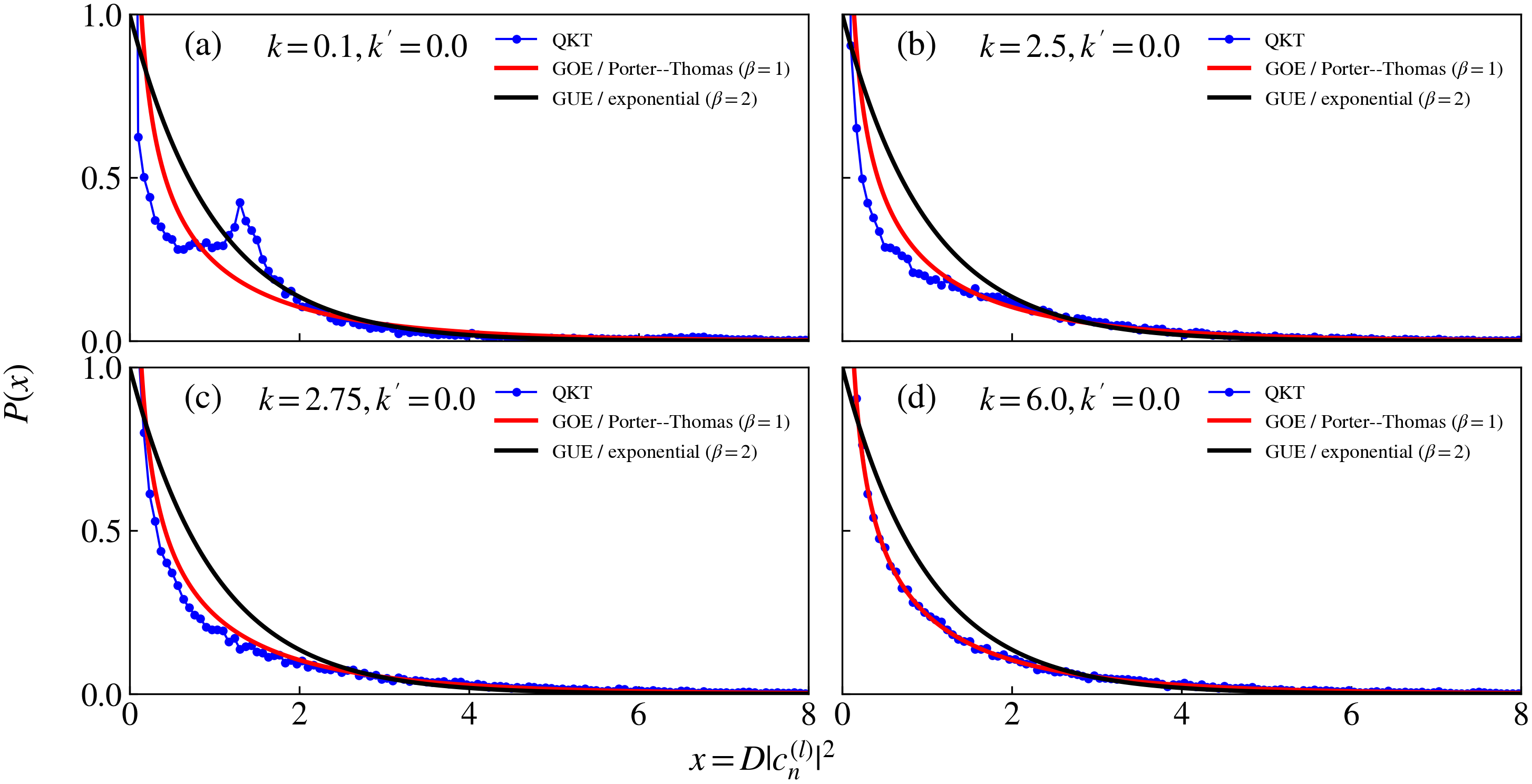}
    \caption{The Porter-Thomus distribution for the QKT with \(p = \pi/2 \) and \(j = 200.5\).}\label{fig:porter-thomus-1}
\end{figure}
\begin{figure}
    \includegraphics[width=\linewidth]{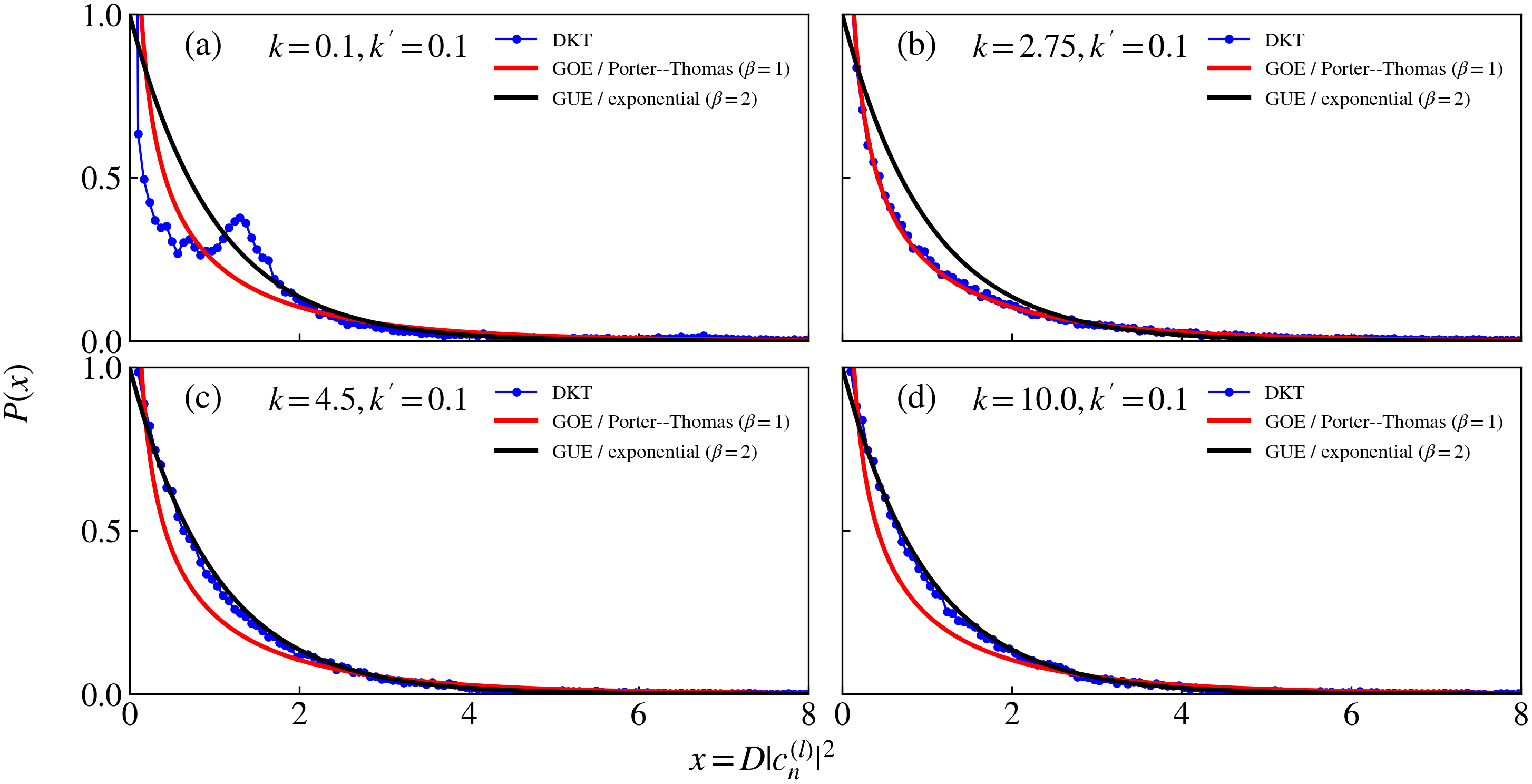}
    \caption{The Porter-Thomus distribution for broken time-reversal-symmetric extension of the QKT given by Eq.~\eqref{eq:dkt} with \(p = \pi/2\) and \(j = 200.5\).}\label{fig:porter-thomus-2}
\end{figure}

Eigenvalue statistics characterize correlations among the eigenphases of the Floquet operator, whereas eigenvector statistics provide complementary information about the structure of individual Floquet eigenstates. For quantum systems whose classical dynamics is chaotic, RMT predicts universal statistics for the components of eigenvectors. This connection was demonstrated explicitly for the QKT by Haake and Życzkowski~\cite{haake1990random, zyczkowski1990indicators, lenz1992time}. They showed that the distribution of eigenvector-components approaches the $\chi^2_\nu$ distributions associated with the orthogonal, unitary, and symplectic universality classes. In particular, QKT in the fully chaotic regime satisfies the Porter--Thomas distribution~\cite{porter1956fluctuations} as it belongs to the orthogonal ($\nu=1$) class (see Figs.~\ref{fig:porter-thomus-1}). While, its broken-time reversal symmetric extension given by~\eqref{eq:dkt} satisfied an exponential distribution as it belongs to the unitary ($\nu=2$) class (see Figs.~\ref{fig:porter-thomus-2} and Ref.~\cite{lenz1992time} for detailed analysis). The coefficients illustrated in Figs.~\ref{fig:porter-thomus-1} and~\ref{fig:porter-thomus-2} are obtained by writing eigenbasis $\lbrace |\phi_l \rangle\rbrace$ in the $\lbrace |j, m \rangle\rbrace$ as follows:
\begin{align}
    |\phi_l \rangle = \sum_{n=1}^{D} c_n^{(l)} |j, m_n \rangle.
\end{align}
Since eigenvector-component statistics are basis-dependent, we use the angular-momentum basis $\{|j,m\rangle\}$, in which $J_z$ is diagonal and the Floquet operator is naturally represented through the angular-momentum operators governing its nonlinear twist and rotation. Similar to the spacing statistics, here also, we can observe that chi-square distribution in the broken time-reversal symmetric case may fit well with the Porter--Thomus. Therefore, in the case of eigenvector statistics also, agreement with a particular RMT distribution should be interpreted together with the symmetry class and the underlying classical phase space structure. 

\subsection{Recurrences}\label{sec:sub:recurrences}
It is not true that quantum systems always mimic their classical counterparts when the system size is large. There exists a set of parameters where quantum systems show behavior that is starkly different from their underlying classical dynamics. Quantum recurrences in the large systems are one such striking manifestation, where the Floquet operator periodically returns arbitrarily close to the initial point after a certain time. In the case of exact recurrences, the system returns exactly to the initial point. Such exact quantum recurrences have also been observed in the QKT models~\cite{anand_quantum_recurrences,purohit2025double}. Since these exact recurrences in the QKT are properties of the Floquet operator (through $\mathcal{U}^n = \mathds{I}$), they do not depend on the initial state. Based on these results, the exact recurrences in the linear entropy for the initial spin-coherent state $|\theta_0 = 0, \phi_0 = 0\rangle$ have been analyzed for 2- to 4-qubit systems in this subsection. Occurrence of exact recurrences for any system size for specific values of $k$ is discussed in brief at the end of this subsection.

\subsubsection{Two-qubit state}
The time-evolved state corresponding to the initial spin-coherent state $|\theta_0 = 0, \phi_0 = 0\rangle$ is given by
\begin{align}
    \begin{split}
        |\psi_n \rangle = \frac{\cos(\frac{n\pi}{2})}{\sqrt{2}}|\Phi_0^+\rangle + \frac{e^{\frac{i k}{4}}\sin(\frac{n\pi}{2})}{\sqrt{2}}|\Phi_1^+ \rangle + \frac{e^{-\frac{i n k}{4}}}{\sqrt{2}} |\Phi_0^-\rangle. 
    \end{split}
\end{align}
By calculating the density matrix and tracing out one qubit, we get the single-qubit RDM. Then, the linear entropy associated with this single-qubit RDM can be computed as follows:
\begin{align}
    S^{(2)}_{(0,0)}\left(n,k\right) = \begin{cases}
        \frac{1}{2}\sin^2\!\left[(n+1)\dfrac{k}{4}\right] & \text{odd } n\\
        \frac{1}{2}\sin^2\!\left(\dfrac{nk}{4}\right) & \text{even } n.
    \end{cases}
\end{align}
From this, one can easily observe that the linear entropy is periodic if $k$ is a rational multiple of $\pi$. Being a highly quantum system, it shows exact recurrences for countably infinite values of $k$.

\subsubsection{Three-qubit state}
The time-evolved state corresponding to the state $|\theta_0 = 0, \phi_0 = 0\rangle$ is given by
\begin{align}
    |\psi_n\rangle = \frac{ \alpha_n}{\sqrt{2}} e^{-\frac{i}{4}n\pi} | \Phi_0^+\rangle 
    + \frac{\beta_n}{\sqrt{2}} e^{-\frac{i}{4}n\pi} | \Phi_1^+\rangle + \frac{\alpha_n}{\sqrt{2}} (-1)^n e^{\frac{i}{4}n\pi} | \Phi_0^-\rangle - \frac{\beta_n}{\sqrt{2}} (-1)^n e^{\frac{i}{4}n\pi} | \Phi_1^-\rangle.
\end{align}
Again, by calculating the density matrix and tracing out over two qubits, one can obtain the linear entropy associated with the single-qubit RDM as follows:
\begin{align}
    S^{(3)}_{(0,0)}(n,k) = \begin{cases}
    \dfrac{4}{9}\,|\beta_n|^2\left(3-2|\beta_n|^2\right) & n\ \text{even} \\
    \dfrac{1}{2}- 2\left(\dfrac{|\beta_n|^2}{3}+\dfrac{\Im(\alpha_n\beta_n^*)}{\sqrt{3}}\right)^{\!2} & n\ \text{odd}.
    \end{cases}
\end{align}
Note that $n$ appears with $\gamma$ through $\alpha_n$, $\beta_n$, their conjugates and products (refer Eqs.~\eqref{Eq:gamma3} and \eqref{Eq:3qubit_alphan}). Both $\alpha_n$ and $\beta_n$ are polynomials in $\gamma$. Therefore, the linear entropy is periodic if there exists a $\gamma = a\pi$ with $a \in \mathds{Q}$ satisfying~\cite{purohit2025double}:
\begin{align}
 \frac{1}{2}\sin\left(\frac{k}{3}\right) = \cos(a\pi) \; \text{and} \; \frac{1}{3} \leq a \leq \frac{2}{3}.
\end{align}
The exact recurrences occur for $k$ satisfying the above condition.

\subsubsection{Four-qubit state}
The time-evolved state corresponding to the state $|\theta_0 = 0, \phi_0 = 0\rangle$ is given by
\begin{align}
    \begin{split}
        |\psi_n\rangle =  e^{-\frac{i}{2}n \left(\frac{k}{2} + \pi\right)} \left[\frac{\alpha'_n}{\sqrt{2}} |\Phi_0^+\rangle + \frac{i\beta'_n}{\sqrt{2}}  |\Phi_2^+\rangle + \frac{\cos(\frac{n\pi}{2})}{\sqrt{2}} e^{-\frac{i}{8}n k}|\Phi_0^-\rangle + \frac{\sin(\frac{n\pi}{2}) e^{-\frac{ik}{8} (n - 3)}}{\sqrt{2}} |\Phi_1^-\rangle \right].
    \end{split}
\end{align}
Again, following a similar analysis, one can find the linear entropy associated with the single-qubit RDM as follows:
\begin{align}
    S^{(4)}_{(0,0)} \left(n, k\right) = \frac{1}{2} - \frac{1+\cos(n\pi)}{4} {\Re[\alpha'_n \delta_n^*]}^2 + \frac{-1+\cos(n\pi)}{16} {\Re[\varepsilon_n^* (\alpha'_n + i\sqrt{3}\beta'_n )]}^2,
\end{align}
where,
\begin{align}
    \Re[\alpha'_n \delta_n^*] =& \cos\left(\frac{nk}{8}\right) \cos(n\gamma) - \frac{1}{2}\frac{\sin(n\gamma)}{\sin(\gamma)} \sin\left(\frac{nk}{8}\right) \cos\left(\frac{k}{2}\right) \\
    \Re[\varepsilon_n^* (\alpha'_n + i\sqrt{3}\beta'_n )] =& - \cos\left( (n-3)\frac{k}{8}\right) \left[\frac{3}{2}\sin\left(\frac{k}{2}\right) \frac{\sin(n\gamma)}{\sin(\gamma)} - \cos(n\gamma) \right]  \notag \\
    & - 2\frac{\sin(n\gamma)}{\sin(\gamma)} \cos\left(\frac{k}{2}\right) \sin\left( (n-3)\frac{k}{8}\right).
\end{align}
The linear entropy is periodic if $\gamma$ and $k$ both are rational multiples of $\pi$. Therefore, the linear entropy is periodic if there exist $a, b \in \mathds{Q}$, such that
\begin{align}
    \begin{split}
        \frac{1}{2}\sin\left(\frac{k}{2}\right) = \cos(a\pi), \; k = b\pi \;\text{ and }\; \frac{1}{3} \leq a \leq \frac{2}{3}.
    \end{split}
\end{align}
The above condition is satisfied for $k = m\pi$, where $m\in\mathds{Z}$. By looking at these exactly solvable systems, we can also observe that such recurrences rapidly become rarer as the system size increases.

\subsubsection{Large systems}
Quantum recurrences for $k>2$ are rare but not absent in large systems. Such exact recurrences have been shown for the system with $k=2\pi j$ and $k=\pi j$~\cite{anand_quantum_recurrences}. For even-qubit systems, exact recurrences were also observed numerically for $k=j\pi/2$ but were absent in odd-qubit systems. The reason for their absence for $k = j\pi/2$ in odd-qubit systems is still unknown. Interestingly, the classical limit does not exist for these values of $k$. Therefore, in spite of the large system size, the QKT operates fully in the quantum regime at these kicking strengths and does not offer any classical interpretation. The QKT shows signatures of integrability for these values, and the transition from integrable to non-integrable becomes sharper with system size. It will be interesting if one can analytically show recurrences other than the exact recurrences and check their initial state dependence.

\section{Signatures of Quantum Chaos}
Several signatures of quantum chaos have been developed over the years. The choice of quantum signature depends on the model and system under consideration. For example, in few-qubit systems, entanglement has been widely used. In large systems, eigenvalue statistics and spectral form factor (SFF) are preferred. The signatures such as the OTOC quickly saturate due to the finite size of Hilbert spaces and a few degrees of freedom. Therefore, they are generally not preferred in the QKT models. In previous sections, entanglement and level-spacing statistics have been used to provide complementary information, with entanglement offering a quantitative measure of quantum correlations and level-spacing statistics characterizing the statistical properties of the spectrum. This section, therefore, aims to provide a brief overview of other widely used signatures of quantum chaos. Each of them measures a certain aspect, such as localization, thermalization, or ergodicity. These have been roughly categorized into three groups: eigenvalue statistics, eigenvector statistics, and dynamical signatures.

\subsection{Eigenvalue statistics}
Among various signatures based on eigenvalues of the Floquet operator (or Hamiltonian operator), widely used techniques are nearest-neighbor level spacing statistics, spacing ratio statistics, and SFF. Among these, the nearest-neighbor spacing has already been discussed in Sec.~\ref{sec:sub:sub:spacing}. Therefore, only the other two techniques are discussed here.

\subsubsection{Spacing ratio statistics}
The ratio statistics are qualitatively similar to the nearest-neighbor statistics, with the advantage that they do not require unfolding of spacings. In this technique, after getting the spacings and sorting them in descending order, ratios are calculated as follows:
\begin{align}
    r_i = \frac{s_{i+1}}{s_i}.
\end{align}
For uncorrelated levels, corresponding to Poisson statistics, the ratio distribution is given by 
\begin{align}
    P_P (r) = \frac{2}{{(1+r)}^2}
\end{align}
with $\langle r\rangle_{\mathrm{P}}=2\ln 2-1\approx0.3863$. In contrast, chaotic systems exhibit Wigner-Dyson statistics described by the standard random-matrix ensembles~\cite{atas2013distribution}. The corresponding RMT distributions for $\beta=1,2,$ and $4$, associated with GOE, GUE, and GSE, respectively, are given by
\begin{align}
    P(r, \beta ) = \frac{1}{Z_\beta} \frac{{\left(r + r^2\right)}^\beta}{{\left(1 + r + r^2\right)}^{1 + \frac{3}{2}\beta}}, \quad \text{where} \quad Z_1 = \frac{8}{27}, \quad Z_2 = \frac{4\pi}{81\sqrt{3}}, \quad Z_4 = \frac{4\pi}{729\sqrt{3}}.
\end{align}
The corresponding mean ratio values are $\langle r\rangle_{\rm GOE}=0.5359$, $\langle r\rangle_{\rm GUE}=0.6027$, and $\langle r\rangle_{\rm GSE}=0.6762$~(see Ref.\cite{atas2013distribution}). It should be noted that all symmetries need to be resolved for studying spacing-ratio statistics. Otherwise, it can give misleading results.

\subsubsection{Spectral form factor}
The SFF is the Fourier transform of the two-point eigenvalue correlation. It grew out of the RMT from the works of~\cite{dyson1970correlations,mehta2004random} and brought into quantum chaos~\cite{berry1985semiclassical}. It probes correlations between eigenvalues at all scales and not just nearest-neighbors. The SFF is defined as follows:
\begin{align}
    K(t) = \left\langle {\left|\tr \mathcal{U}^t\right|}^2 \right\rangle = \left\langle \sum_{l,m} e^{i (\theta_l - \theta_m) t} \right\rangle, \quad t\in \mathds{Z},
\end{align}
where $\langle \cdots \rangle$ indicates average over an ensemble or many realizations. It can be computed using the following steps:
\begin{enumerate}
    \item Diagonalize the Floquet operator $\mathcal{U}$: $\left\lbrace e^{i \theta_l} \right\rbrace$.
    \item Compute $\sum_{l,m} e^{i (\theta_l - \theta_m) t}$ for every realization.
    \item Average: if you have many realizations, then average the SFF over realizations at every time $t$ and find $K(t)$. If you have a single realization, then apply spectral-window or smoothing in time by making sure that smoothing does not wash out the ramp. 
    \item Researchers also use connected-SFF $K_{c}(t)$, which is obtained by subtracting the disconnected part from $K(t)$~\cite{chen2018universal}:
        \begin{align}
            K_{c}(t) = \left\langle {\left|\tr \mathcal{U}^t\right|}^2 \right\rangle  - {\left| \left\langle \tr \mathcal{U}^t \right\rangle \right|}^2 .
        \end{align}
    \item Normalize the SFF to $K(t)/N$ to match the RMT expectations, where $N$ represents the Hilbert space dimension.
    \item Plot ensemble averaged SFF $K(t)/N$ versus time $t$.
\end{enumerate}
Unlike nearest-neighbor spacing statistics, the SFF probes spectral correlations over a broad range of scales and provides information beyond local level repulsion. However, the SFF of an individual finite quantum system is generally not self-averaging. Suitable smoothing or averaging is, therefore, needed for a reliable comparison with RMT~\cite{braun2015self}. This makes the SFF more demanding than local spacing statistics, particularly for finite-dimensional Floquet systems such as the QKT. This may partly explain its limited use in QKT studies. Nevertheless, Braun and Haake demonstrated its usefulness for the fully chaotic QKT after appropriate smoothing~\cite{braun2015self}. For non-unitary dynamics, Li et al.~have also analyzed the dissipative SFF for the dissipative QKT~\cite{li2021spectral}.

\subsection{Eigenvector statistics}
While eigenvalue statistics characterize correlations among the eigenphases of the Floquet operator, eigenvector statistics provide complementary information about the structure of its Floquet eigenstates. In the QKT, eigenvector properties have been studied through the statistics of eigenvector components, participation measures, and phase space localization measures~\cite{haake1990random,zyczkowski1990indicators,sieberer2019digital,wang2023statistics}. These measures provide state-resolved information about localization, delocalization, and ergodicity that is not captured by eigenvalue statistics alone. In particular, in the strongly chaotic regime, the statistics of eigenvector components approach the corresponding RMT predictions, while deviations from these distributions provide information about the regular or mixed character of the dynamics~\cite{haake1990random,zyczkowski1990indicators}.

\subsubsection{Inverse participation ratio}
The IPR quantifies the localization or delocalization of a normalized eigenstate in a given basis. In the purely precessing limit of the QKT, $k=0$, the Floquet operator is diagonal in the $J_y$ basis. Consequently, each Floquet eigenstate is fully localized in this basis, with $\mathrm{IPR}=1$. For $k\neq0$, the nonlinear $J_z^2$ kick is non-diagonal in the $J_y$ basis and causes the Floquet eigenstates to spread over multiple $J_y$ basis states. This is why we choose $J_y$ basis. The IPR and related participation measures have been used to characterize eigenvector delocalization and the onset of quantum chaos in the QKT, including comparisons with RMT predictions~\cite{sieberer2019digital,wang2023statistics}. For an eigenstate $\ket{\psi_\alpha}$ expressed in basis states $\ket{i}$ as $\ket{\psi_\alpha}=\sum_i c_i^{(\alpha)}\ket{i}$, the IPR is defined as follows:
\begin{align}
    \mathrm{IPR}_\alpha=\sum_i\left|c_i^{(\alpha)}\right|^4.
\end{align}
A smaller IPR value implies that the state is widely spread or delocalized across many basis states, whereas a larger IPR indicates localization in a smaller number of basis components. The IPR is inversely related to the effective number of participating basis states and is therefore often expressed through the participation ratio (PR):
\begin{align}
    \mathrm{PR}_\alpha=\frac{1}{\mathrm{IPR}_\alpha}.
\end{align}
The PR provides an estimate of the number of significantly occupied basis states. In a strongly chaotic regime, Floquet eigenstates are expected to become delocalized in a physically appropriate basis and their PR scales with the Hilbert space dimension, $\mathrm{PR}\propto N$, with the proportionality factor depending on the relevant RMT ensemble and basis~\cite{sieberer2019digital}. In regular or localized regimes, the number of significantly occupied basis states can remain much smaller than $N$, resulting in a slower decrease of the IPR with increasing system size. The basis choice is important: the IPR is not a basis-independent quantity, and a basis associated with an integrable or regular limit is particularly useful for diagnosing the spreading induced by the nonlinear kick. For the QKT, the $J_y$ basis is natural because the $k=0$ Floquet operator is diagonal in this basis, so the increase in PR with $k$ directly reflects the mixing of the unperturbed $J_y$ states by the nonlinear $J_z^2$ kick~\cite{sieberer2019digital}.

\subsubsection{Eigenvector-component statistics}
A more direct characterization of RMT-like eigenvectors is obtained from the statistical distribution of their components. Let $\ket{\psi_\alpha}=\sum_i c_i^{(\alpha)}\ket{i}$ be a Floquet eigenstate and define the rescaled eigenvector intensity as $x_i=N|c_i^{(\alpha)}|^2$, where $N$ is the Hilbert space dimension. For a fully chaotic system, the components of the eigenvectors are expected to behave as random variables whose distribution is determined by the corresponding RMT ensemble. In particular, the distribution of $x_i$ approaches a $\chi^2_\nu$ distribution, with $\nu=1$, $2$, and $4$ for the orthogonal, unitary, and symplectic ensembles, respectively~\cite{haake1990random,zyczkowski1990indicators}. For the orthogonal ensemble, $\nu=1$ gives the Porter--Thomas distribution. These eigenvector statistics were explicitly demonstrated for the kicked top, where the eigenvector-component distribution approaches the RMT prediction in the strongly chaotic regime and deviates from it in the regular and mixed regimes~\cite{haake1990random}. Thus, unlike the IPR, which reduces the information contained in an eigenvector to a single number, the component distribution provides a direct test of whether the full eigenvector statistics are consistent with the appropriate RMT ensemble.

\subsubsection{Phase space localization and multifractality}
The basis dependence of the IPR can be reduced by examining the localization of Floquet eigenstates directly in the classical phase space. For the QKT, a natural representation is provided by the Husimi function, which is obtained by projecting a Floquet eigenstate onto spin-coherent states $|\theta,\phi\rangle$. For an eigenstate $\ket{\psi_\alpha}$, the Husimi function is defined as follows:
\begin{align}
    Q_\alpha(\theta,\phi)=\left|\langle\theta,\phi|\psi_\alpha\rangle\right|^2.
\end{align}
The Husimi function provides a positive phase space representation of the eigenstate and allows its localization to be compared directly with the underlying classical phase space structures. In regular regions, Floquet eigenstates tend to remain localized around the corresponding classical invariant structures, whereas in strongly chaotic regions they become broadly distributed over the accessible phase space~\cite{zyczkowski1990indicators,wang2023statistics}.

The localization of the Husimi function can be quantified using measures such as the phase space inverse participation ratio and the Wehrl entropy. The latter is defined as follows:
\begin{align}
    S_{\mathrm{W}}^{(\alpha)}=-\int d\Omega\,Q_\alpha(\theta,\phi)\ln Q_\alpha(\theta,\phi),
\end{align}
where $d\Omega$ denotes the appropriate measure on the Bloch sphere. A localized eigenstate has a smaller Wehrl entropy, whereas a delocalized eigenstate has a larger Wehrl entropy. Similarly, the phase space IPR provides a measure of the inverse effective area occupied by the Husimi distribution. These measures have been used to characterize the transition from regular to chaotic dynamics in the QKT and to compare quantum eigenstate localization with the underlying classical phase space structure~\cite{wang2023statistics}.

A related approach studies the expansion of spin-coherent states in the Floquet-eigenstate basis, $|\theta,\phi\rangle=\sum_\alpha c_\alpha(\theta,\phi)|\psi_\alpha\rangle$. The statistical properties of the coefficients $|c_\alpha|^2$ can reveal how a coherent state explores the quasienergy spectrum. In the regular regime, the expansion is concentrated on a relatively small number of Floquet eigenstates, whereas in the chaotic regime it becomes broadly distributed. The scaling of generalized participation ratios with Hilbert space dimension can further reveal multifractal behavior in the transition region between regular and chaotic dynamics~\cite{wang2021multifractality}. Thus, phase space localization and multifractal measures provide a bridge between eigenvector statistics in Hilbert space and the classical structures represented in phase space.

\subsubsection{Eigenstate entanglement entropy}
The EEE is a static property that characterizes the intrinsic entanglement structure of individual Floquet eigenstates. Unlike the dynamical entanglement entropy discussed earlier, which characterizes the entanglement generated during time evolution, the EEE characterizes the entanglement content of individual eigenstates of the Floquet operator. It has been investigated in coupled kicked tops to characterize the crossover toward RMT behavior~\cite{harrmann2020entanglement}. Given a Floquet eigenstate $\ket{\psi_\alpha}$, one can partition the constituent spin-$1/2$ degrees of freedom into subsystems $A$ and $B$ and define the reduced density matrix $\rho_A=\tr_B\left(\ket{\psi_\alpha}\bra{\psi_\alpha}\right)$. The EEE is then given by
\begin{align}
    S_A^{(\alpha)}=-\tr\left(\rho_A\ln\rho_A\right).
\end{align}
In chaotic regimes, Floquet eigenstates generally exhibit stronger entanglement and their entanglement properties can approach random-state or RMT expectations, whereas eigenstates associated with regular structures tend to exhibit lower entanglement~\cite{harrmann2020entanglement}. Thus, EEE provides a complementary eigenvector-based diagnostic: while the IPR measures localization in a chosen basis, eigenvector-component statistics test random-vector behavior, and Husimi-based measures quantify localization in classical phase space, the EEE probes quantum correlations between subsystems. Together, these measures provide complementary information about the structure of Floquet eigenstates and their connection to the underlying classical dynamics.

\subsection{Dynamical signatures}
Except for the SFF, other diagnoses discussed so far in this section are based on static properties of eigenvalues and eigenvectors. Although the SFF depends on time, it crucially relies on the spectral properties of the Floquet operator (or Hamiltonian). Therefore, it has been categorized under eigenvalue statistics. Among various dynamical signatures, a few widely used techniques include the dynamical entropy, OTOC, and Loschmidt echo (LE). Since the dynamical entropy has already been discussed in Sec.~\ref{sec:quantum}, the remaining two techniques will now be discussed.

\subsubsection{Out-of-time-ordered correlators}
The OTOC was first introduced by Larkin and Ovchinnikov in the context of quasi-classical superconductivity~\cite{larkin1969quasiclassical}, and later re-emerged as a central diagnostic of quantum chaos and information scrambling in many-body systems, notably through work by Kitaev and subsequent developments in the SYK model and holography~\cite{kobrin2021many, maldacena2016bound}. OTOCs are now widely used to connect operator growth with classical Lyapunov behavior~\cite{rozenbaum2017lyapunov, trunin2023refined, majewski2006quantum, falsaperla2002quantum} and to characterize information scrambling in many-body quantum systems~\cite{kobrin2021many}, including systems described by quantum field theories~\cite{maldacena2016bound}.

For two Heisenberg operators $\hat{W}(t)$ and $\hat{V}(0)$, a commonly used diagnostic is the squared commutator given by~\cite{larkin1969quasiclassical, maldacena2016bound}
\begin{align}
    C(t) = \left\langle {[\hat{W}(t),\hat{V}(0)]}^{\dagger} {[\hat{W}(t),\hat{V}(0)]} \right\rangle = - \tr\left(\rho\, [\hat{W}(t),\hat{V}(0)]^{2} \right) \; \text{and} \; \rho = \frac{1}{2j+1}\mathds{I}.
\end{align}
The second equality in above equation requires suitable Hermiticity assumptions. The quantity $C(t)$ quantifies how strongly the two operators fail to commute as time evolves. This quantity is closely related to the following one:
\begin{align}
    F(t) = \langle \hat{W}^\dagger(t)\,\hat{V}^\dagger(0)\,\hat{W}(t)\,\hat{V}(0)\rangle .
\end{align}
The expectation values can be taken in a state such as an eigenstate or with respect to a thermal density matrix. As time increases, the growth of the commutator $C(t)$ is reflected in a corresponding decay of $F(t)$. If initially localized operators spread rapidly, making $C(t) \sim e^{2\lambda_Q t}$, where $\lambda_Q$ is the quantum Lyapunov exponent. Then, the system is said to be quantum chaotic. 

For finite many-body systems, OTOCs are evaluated numerically as follows:
\begin{enumerate}
    \item Choose operators $\hat{W}$ and $\hat{V}$, for example local spin operators at different sites. This choice is crucial in diagnosing quantum chaos.
    \item Diagonalize the Hamiltonian and construct the time-evolved operator $\hat{W}(t) = e^{i\hat{H}t}\hat{W}e^{-i\hat{H}t}$.
    \item Obtain the operator $C(t)$ as a trace with a density matrix $\rho$ or as an expectation value in a given eigenstate.
\end{enumerate}
For larger systems where exact diagonalization is not feasible, tensor-network approaches are used to approximate the Heisenberg evolution and compute OTOCs.

\subsubsection{Loschmidt Echo}
The LE was introduced by Peres as a measure of the stability of quantum evolution under small perturbations of the Hamiltonian~\cite{peres1984stability}. Since then, it has become a standard dynamical probe of quantum chaos and decoherence, with extensive semiclassical and numerical studies~\cite{cucchietti2003decoherence,gorin2006dynamics}. In systems with a well-defined classical limit, the decay of the LE can exhibit a Lyapunov regime in which the decay rate is governed by the classical Lyapunov exponent, providing a direct dynamical signature of chaos in the quantum regime~\cite{jalabert2001environment,cucchietti2003universality}. The decay can also occur in a perturbative or Fermi-golden-rule regime, where the decay rate depends on the perturbation strength~\cite{jacquod2001golden}.

Given an initial state $|\psi_0\rangle$, one defines forward evolution under a ``reference'' Hamiltonian $\hat{H}_0$ and backward evolution under a slightly perturbed Hamiltonian $\hat{H}=\hat{H}_0+\hat{V}$. The LE at time $t$ is then given by
\begin{align}
    M(t) = \left|\langle\psi_0|e^{i\hat{H}t}e^{-i\hat{H}_0 t}|\psi_0\rangle\right|^{2}.
\end{align}
It quantifies the sensitivity of the quantum evolution to the perturbation and can be interpreted as a measure of irreversibility. In quantum chaotic systems, the LE decays as $M(t)\sim e^{-\Gamma t}$, with a rate $\Gamma$ depending on the perturbation strength. In systems with a well-defined classical limit and  window, this decay rate can be governed by the Lyapunov exponent. Regular systems typically show non-exponential decay or oscillatory behavior.

\section{Experimental realizations}
Experimental realizations not only make QKT a paradigmatic model but also offer insights into fundamental questions about thermalization, ergodicity, and information scrambling in isolated quantum systems. By implementing the QKT on controllable quantum platforms, researchers have been able to visualize phase space structures, measure entanglement growth, thermalization, and the impact of chaos on quantum sensing. The following subsections briefly discuss three such experimental realizations of the QKT.

\subsection{Quantum signatures of chaos in a kicked top}
The QKT was experimentally realized using the collective spin of a single $^{133}$Cs atom in the angular momentum $F=3$ hyperfine ground state~\cite{chaudhary_quantum_signatures}. In this experiment, the nonlinear kick was generated by the AC Stark shift from a linearly polarized laser field tuned near the D$_1$ resonance at $895\,\mathrm{nm}$, while the linear precession was implemented by applying short magnetic-field pulses along the $y$-axis. Initial states were prepared as spin-coherent states $|\theta,\phi\rangle$ by optical pumping and controlled magnetic-field rotations. After each kick, the quantum state was reconstructed using continuous weak optical measurements combined with dynamical control, allowing the evolution of the quantum phase space distribution to be directly observed~\cite{chaudhary_quantum_signatures}.

The reconstructed states were visualized using the Husimi $Q$-function, which revealed a clear correspondence between quantum dynamics and the underlying classical phase space structures. States initialized in regular regions remained localized, whereas those initialized in chaotic regions rapidly spread over the accessible phase space. Dynamical tunnelling between classically disconnected regular islands was also observed. In addition, the experiment demonstrated differences in the sensitivity to perturbations between regular and chaotic regimes and provided experimental evidence for dynamical entanglement as a signature of chaos~\cite{chaudhary_quantum_signatures}. Thus, the atomic-spin realization provided direct experimental access to quantum--classical correspondence and to dynamical signatures of chaos in the QKT.

\subsection{Ergodic dynamics and thermalization in an isolated quantum system}
Ergodic dynamics and thermalization in the quantum kicked top were experimentally investigated~\cite{neil2016ergodic} using a superconducting quantum processor based on three fixed-frequency transmon qubits. Rotations around the $y$-axis were performed using shaped microwave pulses, while the nonlinear interaction $J_z^2$ was implemented using a tunable coupling circuit controlled by three separate square pulses. The kicking strength $\kappa=3gT/\hbar$ was determined from the qubit-qubit interaction energy $g$ and the duration $T$ of the interaction pulses. Initial states were prepared as spin-coherent states, and the unitary operations were repeated for $N$ time steps before measurement. The evolution of the quantum state was characterized through single- and three-qubit state tomography, allowing both local and global properties of the evolving state to be investigated.

The primary signature of chaos was revealed through measurements of entanglement entropy. The time-averaged von Neumann entropy of a single-qubit reduced density matrix showed a strong correspondence with the underlying classical phase space structure. Initial states associated with classically chaotic regions generated higher entanglement, whereas states in regular regions exhibited lower entanglement. The authors further showed that regions of high entanglement were associated with ergodic dynamics and that the entropy approached its maximum value for stronger interactions.

\subsection{Quantum metrology with quantum-chaotic sensors}
Fiderer and Braun used the QKT as a ``quantum--chaotic sensor'' by embedding its dynamics into realistic spin–precession magnetometers~\cite{fiderer2018quantum}. Here, the sensor was an effective large spin that precessed in the magnetic field to be estimated, while periodic nonlinear kicks made the dynamics chaotic and increased the quantum Fisher information (QFI) without requiring highly entangled input states. The QFI was evaluated via LE. In the idealized model, a single large pseudo–spin undergoes regular Larmor precession about a fixed axis and gets interrupted by strong, parameter–independent nonlinear pulses that implement an effective quadratic spin term.

To model realistic dissipation, the QKT dynamics were described by the Markovian master equation as follows:
\begin{align}\label{eq:markov}
  \frac{d\rho}{dt} = 2\gamma\Bigl( J_- \rho J_+ - \tfrac{1}{2}\{J_+ J_-,\rho\}\Bigr),
\end{align}
where $\gamma$ is the dissipation rate and $J_\pm$ are collective ladder operators. Since this dissipation operator commutes with the precession Hamiltonian, the evolution of the density operator can be given by
\begin{align}
  \rho_{t+1} = U_k\,\mathcal{D}_\gamma(\rho_t)\,U_k^\dagger,
\end{align}
where $U_k$ is the unitary Floquet operator and $\mathcal{D}_\gamma$ the channel generated by Eq.~\eqref{eq:markov} over one period. In this open setting, kicks create non–equilibrium steady states with finite QFI. This is in contrast to the non–kicked case, where the top relaxes to a ground state.

This implementation shows that the QKT model maps naturally onto alkali–vapor magnetometers. When the sensor is driven into a chaotic regime, it leads to a significant QFI enhancement over an integrable precession sensor with the same resources. This was observed in both closed and dissipative settings. Furthermore, when implemented in realistic cesium SERF magnetometers, the protocol gave a substantial gain in rescaled QFI and improved magnetic field sensitivity without relying on fragile entangled states.

\section{Discussion and Outlook}
In this chapter, the QKT model has been presented as a unifying framework for exploring nonlinear dynamics, quantum chaos, and experimental realizations of chaos. Its simplicity and versatility make it an ideal testbed for studying the interplay between classical chaos and quantum information.

In classical dynamics, the derivation of the stroboscopic map, fixed point analysis, and computation of the LLE show how such a simple model can generate a full range of dynamics from regular to fully chaotic behavior. The model can be represented as a collection of spin-1/2 particles with two-body interaction. This consideration exhibits rich entanglement dynamics, spectral statistics, quantum recurrences and also encodes signatures of the underlying phase space in the semiclassical limit. In the broader literature, the QKT has long been recognized as a benchmark model for studying Hamiltonian chaos and classical-quantum correspondence.

Pedagogically, the chapter prepares the reader for advanced studies along several directions. By working through the construction of the Floquet operator and the associated classical map, the reader gains hands-on experience with stroboscopic dynamics. The fixed-point analysis, supported by phase space portraits and LLE landscapes, trains the reader to recognize and quantify how chaos develops in the system. The quantum treatment views the kicked top as an ensemble of interacting spin degrees of freedom and introduces tools important for quantum information. These include RDMs, quantum correlations, statistical measures, and tools that reveal thermalization or lack thereof.

Having surveyed the existing literature, the QKT also points toward several open research directions. Each signature of quantum chaos has a domain of validity depending on its observable properties, such as delocalization, thermalization, or information scrambling. Yet, whether any of these unambiguously imply classical chaos remains an open question. The relationships among different signatures continue to be investigated. While classical dynamics clearly distinguishes between ergodicity and chaos, the distinction becomes less defined in quantum systems. Whether quantum indicators can distinguish ergodicity and chaos is still unresolved. In certain regimes, quantum interference can suppress classically chaotic diffusion, giving rise to effects (see discussion in Sec.~\ref{sec:sub:high-spin}) such as dynamical localization similar to the kicked rotor~\cite{fishman1982chaos, tian2004weak}. Understanding the mechanisms and conditions under which such suppression occurs is still under investigation.

The stability analysis in this chapter focuses on local fixed points and their linear stability. However, global bifurcations require topological and nonlinear tools that are beyond linear analysis. This raises further questions: Like local bifurcations, do global bifurcations leave any signature in quantum observables? Is there a minimum scale such as the Planck scale or finite spin-size below which quantum dynamics cannot resolve topological structures? Which topological aspects are universal across QKT systems, and which depend on the specific details of the model?

Finally, extensions of the model suggest a broader research program at the interface of quantum chaos and quantum technologies. Kicked $m$-spin models generalize the QKT to include many-body interactions, enabling new avenues for investigation. Similarly, incorporating dissipation and realistic noise channels into QKT dynamics connects the abstract theory of quantum chaos with practical issues such as error growth, system robustness, and the reliability of quantum simulators and sensors.

\bibliographystyle{JHEP}
\bibliography{ref}

\providecommand{\href}[2]{#2}\begingroup\raggedright\begin{thebibliography}{10}

\bibitem{haake1987classical}
F.~Haake, M.~Ku{\'s} and R.~Scharf, \emph{Classical and quantum chaos for a kicked top}, {\emph{Zeitschrift f{\"u}r Physik B Condensed Matter} {\bfseries 65} (1987) 381}.

\bibitem{casati1980connection}
G.~Casati, F.~Valz-Gris and I.~Guarnieri, \emph{On the connection between quantization of nonintegrable systems and statistical theory of spectra}, {\emph{Lettere al Nuovo Cimento (1971-1985)} {\bfseries 28} (1980) 279}.

\bibitem{Izrailev1989}
F.M.~Izrailev, \emph{Intermediate statistics of the quasi-energy spectrum and quantum localisation of classical chaos}, {\emph{J. Phys. A: Mathematical and General} {\bfseries 22} (1989) 865}.

\bibitem{chirikov1979universal}
B.V.~Chirikov, \emph{A universal instability of many-dimensional oscillator systems}, {\emph{Physics reports} {\bfseries 52} (1979) 263}.

\bibitem{purohit2025double}
A.V.~Purohit and U.T.~Bhosale, \emph{Study of the double kicked top: A classical and quantum perspective}, {\emph{Phys. Rev. E} {\bfseries 112} (2025) 014217}.

\bibitem{kuznetsov1998elements}
Y.A.~Kuznetsov, \emph{Elements of applied bifurcation theory}, Springer (1998).

\bibitem{guckenheimer2013nonlinear}
J.~Guckenheimer and P.~Holmes, \emph{Nonlinear oscillations, dynamical systems, and bifurcations of vector fields}, Springer Science \& Business Media (2013).

\bibitem{p-spin_poggi}
M.H.~Mu\~noz Arias, P.M.~Poggi and I.H.~Deutsch, \emph{Nonlinear dynamics and quantum chaos of a family of kicked $p$-spin models}, {\emph{Phys. Rev. E} {\bfseries 103} (2021) 052212}.

\bibitem{Bandyopadhyay2004}
J.N.~Bandyopadhyay and A.~Lakshminarayan, \emph{Entanglement production in coupled chaotic systems: Case of the kicked tops}, {\emph{Phys. Rev. E} {\bfseries 69} (2004) 016201}.

\bibitem{sreeram2021out}
P.~Sreeram, V.~Madhok and A.~Lakshminarayan, \emph{Out-of-time-ordered correlators and the loschmidt echo in the quantum kicked top: how low can we go?}, {\emph{J. Phys. D: Applied Physics} {\bfseries 54} (2021) 274004}.

\bibitem{rozenbaum2017lyapunov}
E.B.~Rozenbaum, S.~Ganeshan and V.~Galitski, \emph{Lyapunov exponent and out-of-time-ordered correlator’s growth rate in a chaotic system}, {\emph{Phys. Rev. Lett.} {\bfseries 118} (2017) 086801}.

\bibitem{trunin2023refined}
D.A.~Trunin, \emph{Refined quantum lyapunov exponents from replica out-of-time-order correlators}, {\emph{Phys. Rev. D} {\bfseries 108} (2023) 105023}.

\bibitem{majewski2006quantum}
W.A.~Majewski and M.~Marciniak, \emph{On quantum lyapunov exponents}, {\emph{Journal of Physics A: Mathematical and General} {\bfseries 39} (2006) L523}.

\bibitem{falsaperla2002quantum}
P.~Falsaperla, G.~Fonte and G.~Salesi, \emph{Quantum lyapunov exponents}, {\emph{Foundations of Physics} {\bfseries 32} (2002) 267}.

\bibitem{krithika2019nmr}
V.~Krithika, V.~Anjusha, U.T.~Bhosale and T.~Mahesh, \emph{Nmr studies of quantum chaos in a two-qubit kicked top}, {\emph{Phys. Rev. E} {\bfseries 99} (2019) 032219}.

\bibitem{neil2016ergodic}
C.~Neill, P.~Roushan, M.~Fang, Y.~Chen, M.~Kolodrubetz, Z.~Chen et~al., \emph{Ergodic dynamics and thermalization in an isolated quantum system}, {\emph{Nat. Phys.} {\bfseries 12} (2016) 1037}.

\bibitem{dogra2019quantum}
S.~Dogra, V.~Madhok and A.~Lakshminarayan, \emph{Quantum signatures of chaos, thermalization, and tunneling in the exactly solvable few-body kicked top}, {\emph{Phys. Rev. E} {\bfseries 99} (2019) 062217}.

\bibitem{lerose2020bridging}
A.~Lerose and S.~Pappalardi, \emph{Bridging entanglement dynamics and chaos in semiclassical systems}, {\emph{Phys. Rev. A} {\bfseries 102} (2020) 032404}.

\bibitem{lombardi2011entanglement}
M.~Lombardi and A.~Matzkin, \emph{Entanglement and chaos in the kicked top}, {\emph{Phys. Rev. E} {\bfseries 83} (2011) 016207}.

\bibitem{Wang2004entanglement}
X.~Wang, S.~Ghose, B.C.~Sanders and B.~Hu, \emph{Entanglement as a signature of quantum chaos}, {\emph{Phys. Rev. E} {\bfseries 70} (2004) 016217}.

\bibitem{Bandyopadhyay2004entanglement}
J.N.~Bandyopadhyay and A.~Lakshminarayan, \emph{Entanglement production in coupled chaotic systems: Case of the kicked tops}, {\emph{Phys. Rev. E} {\bfseries 69} (2004) 016201}.

\bibitem{Bianchi2018}
E.~Bianchi, L.~Hackl and N.~Yokomizo, \emph{Linear growth of the entanglement entropy and the kolmogorov-sinai rate}, {\emph{J. High Energy Phys.} {\bfseries 2018} (2018) 25}.

\bibitem{wigner1967random}
E.P.~Wigner, \emph{Random matrices in physics}, {\emph{SIAM review} {\bfseries 9} (1967) 1}.

\bibitem{bohigas1991random}
O.~Bohigas et~al., \emph{Random matrix theories and chaotic dynamics},  Tech. Rep. Paris-11 Univ., 91-Orsay (France). Inst. de Physique Nucleaire (1991).

\bibitem{mehta2004random}
M.L.~Mehta, \emph{Random Matrices}, Elsevier Academic Press, London, 3rd~ed. (2004).

\bibitem{kus1987symmetry}
M.~Ku{s}, R.~Scharf and F.~Haake, \emph{Symmetry versus degree of level repulsion for kicked quantum systems}, {\emph{Z. Phys. B} {\bfseries 66} (1987) 129}.

\bibitem{haake1990random}
F.~Haake and K.~\ifmmode~\dot{Z}\else \.{Z}\fi{}yczkowski, \emph{Random-matrix theory and eigenmodes of dynamical systems}, {\emph{Phys. Rev. A} {\bfseries 42} (1990) 1013(R)}.

\bibitem{zyczkowski1990indicators}
K.~Zyczkowski, \emph{Indicators of quantum chaos based on eigenvector statistics}, {\emph{Journal of Physics A: Mathematical and General} {\bfseries 23} (1990) 4427}.

\bibitem{lenz1992time}
G.~Lenz and K.~Zyczkowski, \emph{Time-reversal symmetry breaking and the statistical properties of quantum systems}, {\emph{Journal of Physics A: Mathematical and General} {\bfseries 25} (1992) 5539}.

\bibitem{porter1956fluctuations}
C.E.~Porter and R.G.~Thomas, \emph{Fluctuations of nuclear reaction widths}, {\emph{Phys. Rev.} {\bfseries 104} (1956) 483}.

\bibitem{anand_quantum_recurrences}
A.~Anand, J.~Davis and S.~Ghose, \emph{Quantum recurrences in the kicked top}, {\emph{Phys. Rev. Res.} {\bfseries 6} (2024) 023120}.

\bibitem{atas2013distribution}
Y.Y.~Atas, E.~Bogomolny, O.~Giraud and G.~Roux, \emph{Distribution of the ratio of consecutive level spacings in random matrix ensembles}, {\emph{Phys. Rev. Lett.} {\bfseries 110} (2013) 084101}.

\bibitem{dyson1970correlations}
F.J.~Dyson, \emph{Correlations between eigenvalues of a random matrix}, {\emph{Communications in Mathematical Physics} {\bfseries 19} (1970) 235}.

\bibitem{berry1985semiclassical}
M.V.~Berry, \emph{Semiclassical theory of spectral rigidity}, {\emph{Proceedings of the Royal Society of London. A. Mathematical and Physical Sciences} {\bfseries 400} (1985) 229}.

\bibitem{chen2018universal}
X.~Chen and A.W.W.~Ludwig, \emph{Universal spectral correlations in the chaotic wave function and the development of quantum chaos}, {\emph{Phys. Rev. B} {\bfseries 98} (2018) 064309}.

\bibitem{braun2015self}
P.~Braun and F.~Haake, \emph{Self-averaging characteristics of spectral fluctuations}, {\emph{Journal of Physics A: Mathematical and Theoretical} {\bfseries 48} (2015) 135101}.

\bibitem{li2021spectral}
J.~Li, T.c.v.~Prosen and A.~Chan, \emph{Spectral statistics of non-hermitian matrices and dissipative quantum chaos}, {\emph{Phys. Rev. Lett.} {\bfseries 127} (2021) 170602}.

\bibitem{sieberer2019digital}
L.M.~Sieberer, T.~Olsacher, A.~Elben, M.~Heyl, P.~Hauke, F.~Haake et~al., \emph{Digital quantum simulation, trotter errors, and quantum chaos of the kicked top}, {\emph{npj Quantum Information} {\bfseries 5} (2019) 78}.

\bibitem{wang2023statistics}
Q.~Wang and M.~Robnik, \emph{Statistics of phase space localization measures and quantum chaos in the kicked top model}, {\emph{Phys. Rev. E} {\bfseries 107} (2023) 054213}.

\bibitem{wang2021multifractality}
Q.~Wang and M.~Robnik, \emph{Multifractality in quasienergy space of coherent states as a signature of quantum chaos}, {\emph{Entropy} {\bfseries 23} (2021) 1347}.

\bibitem{harrmann2020entanglement}
T.~Herrmann, M.F.I.~Kieler, F.~Fritzsch and A.~B\"acker, \emph{Entanglement in coupled kicked tops with chaotic dynamics}, {\emph{Phys. Rev. E} {\bfseries 101} (2020) 022221}.

\bibitem{larkin1969quasiclassical}
A.I.~Larkin and Y.N.~Ovchinnikov, \emph{Quasiclassical method in the theory of superconductivity}, {\emph{Sov Phys JETP} {\bfseries 28} (1969) 1200}.

\bibitem{kobrin2021many}
B.~Kobrin, Z.~Yang, G.D.~Kahanamoku-Meyer, C.T.~Olund, J.E.~Moore, D.~Stanford et~al., \emph{Many-body chaos in the sachdev-ye-kitaev model}, {\emph{Phys. Rev. Lett.} {\bfseries 126} (2021) 030602}.

\bibitem{maldacena2016bound}
J.~Maldacena, S.H.~Shenker and D.~Stanford, \emph{A bound on chaos}, {\emph{J. High Energy Phys.} {\bfseries 2016} (2016) 1}.

\bibitem{peres1984stability}
A.~Peres, \emph{Stability of quantum motion in chaotic and regular systems}, {\emph{Phys. Rev. A} {\bfseries 30} (1984) 1610}.

\bibitem{cucchietti2003decoherence}
F.M.~Cucchietti, D.A.~Dalvit, J.P.~Paz and W.H.~Zurek, \emph{Decoherence and the loschmidt echo}, {\emph{Phys. Rev. Lett.} {\bfseries 91} (2003) 210403}.

\bibitem{gorin2006dynamics}
T.~Gorin, T.~Prosen, T.H.~Seligman and M.~Znidaric, \emph{Dynamics of loschmidt echoes and fidelity decay}, {\emph{Physics Reports} {\bfseries 435} (2006) 33}.

\bibitem{jalabert2001environment}
R.A.~Jalabert and H.M.~Pastawski, \emph{Environment-independent decoherence rate in classically chaotic systems}, {\emph{Phys. Rev. Lett.} {\bfseries 86} (2001) 2490}.

\bibitem{cucchietti2003universality}
F.M.~Cucchietti, H.M.~Pastawski and R.A.~Jalabert, \emph{Universality of the lyapunov regime for the loschmidt echo}, {\emph{Phys. Rev. B} {\bfseries 70} (2004) 035311}.

\bibitem{jacquod2001golden}
P.~Jacquod, P.~Silvestrov and C.~Beenakker, \emph{Golden rule decay versus lyapunov decay of the quantum loschmidt echo}, {\emph{Phys. Rev. E} {\bfseries 64} (2001) 055203(R)}.

\bibitem{chaudhary_quantum_signatures}
S.~Chaudhury, A.~Smith, B.~Anderson, S.~Ghose and P.S.~Jessen, \emph{Quantum signatures of chaos in a kicked top}, {\emph{Nature} {\bfseries 461} (2009) 768}.

\bibitem{fiderer2018quantum}
L.J.~Fiderer and D.~Braun, \emph{Quantum metrology with quantum-chaotic sensors}, {\emph{Nat. Commun.} {\bfseries 9} (2018) 1351}.

\bibitem{fishman1982chaos}
S.~Fishman, D.R.~Grempel and R.E.~Prange, \emph{Chaos, quantum recurrences, and anderson localization}, {\emph{Phys. Rev. Lett.} {\bfseries 49} (1982) 509}.

\bibitem{tian2004weak}
C.~Tian, A.~Kamenev and A.~Larkin, \emph{Weak dynamical localization in periodically kicked cold atomic gases}, {\emph{Phys. Rev. Lett.} {\bfseries 93} (2004) 124101}.

\end{thebibliography}\endgroup

\appendix

\section{Classical Map}\label{appendix:classical_map}
Evolution of an initial angular momentum vector $\mathbf{J} = (J_x, J_y, J_z)$ to $\mathbf{J}'$ after one period $\tau$ is given by
\begin{align}
    \mathbf{J}' = \exp\left(i p J_y\right) \exp\left(i \frac{k}{2j}J_z^2\right) \mathbf{J} \exp\left(- i \frac{k}{2j}J_z^2\right) \exp\left(- i p J_y\right).
\end{align}
Since the classical limit is obtained by taking $\mathbf{X} = \lim_{j\to\infty} \frac{\mathbf{J}}{j}$, we simplify the transformation into three steps before taking the classical limit:

(1) Nonlinear part: $\exp\left(i \frac{k}{2j}J_z^2\right) \mathbf{J} \exp\left(- i \frac{k}{2j}J_z^2\right)$. 

(2) Linear part: $\exp\left(i p J_y\right) \mathbf{J} \exp\left(- i p J_y\right)$.

(3) Merge the first two parts together using identities to get $\mathcal{U}^\dagger\mathbf{J}\mathcal{U}$.

\textit{Nonlinear part:} Since the states $|j,m\rangle$ are eigenstates of the $J_z$ operator, the operator $\exp\left(-i \frac{k}{2j}J_z^2\right)$ is diagonal in these bases. Furthermore, $J_x$ and $J_y$ can be expressed in terms of raising and lowering operators $J_\pm = J_x \pm i J_y$. Then, the action of the nonlinear part on $J_\pm$ is given by
\begin{align}
    \langle j, m | e^{i \frac{k}{2j}J_z^2} J_\pm e^{-i \frac{k}{2j}J_z^2}  | j, n  \rangle = e^{i \frac{k}{2j} \left(m^2 - n^2\right)} C_{j m, n \pm 1 }\delta_{m, n \pm 1}.
\end{align}
The right-hand side is non-zero only when $m = n \pm 1$ respectively. Therefore, setting $m^2 = n^2 \pm 2n + 1$, we get 
\begin{align}
    \langle j, n \pm 1 | e^{i \frac{k}{2j}J_z^2} J_\pm e^{-i \frac{k}{2j}J_z^2} | j, n  \rangle = e^{i \frac{k}{j} \left(\pm n - \frac{1}{2}\right)} C_{j n \pm 1 } 
    \implies  e^{i \frac{k}{2j}J_z^2} J_\pm e^{-i \frac{k}{2j}J_z^2} = e^{i \frac{k}{j} \left(\pm J_z - \frac{1}{2}\right)} J_\pm .
\end{align}
Since $J_x = \frac{1}{2}(J_+ + J_-)$, we get
\begin{align}\label{Eq:nonlinear:Jx}
    e^{i \frac{k}{2j}J_z^2} J_x e^{-i \frac{k}{2j}J_z^2}  &= \frac{1}{2} \left[ e^{i \frac{k}{j} \left( J_z -\frac{1}{2}\right)} J_+ + e^{ -i \frac{k}{j} \left( J_z + \frac{1}{2}\right)} J_- \right] \notag \\
    &= \frac{e^{-i \frac{k}{2j}}}{2} \left[ e^{i \frac{k}{j} J_z} J_x + e^{ -i \frac{k}{j} J_z} J_x + i e^{i \frac{k}{j} J_z} J_y - i e^{ -i \frac{k}{j} J_z} J_y \right] \notag \\
    &= e^{-i \frac{k}{2j}} \left[ \cos\left(\frac{k}{j} J_z\right) J_x - \sin\left(\frac{k}{j} J_z\right) J_y \right].
\end{align}
Similarly, the operator $J_y$ is transformed as follows:
\begin{align}\label{Eq:nonlinear:Jy}
    e^{i \frac{k}{2j}J_z^2} J_y e^{-i \frac{k}{2j}J_z^2}  &= \frac{1}{2i} \left[ e^{i \frac{k}{j} \left( J_z -\frac{1}{2}\right)} J_+ - e^{ -i \frac{k}{j} \left( J_z + \frac{1}{2}\right)} J_- \right] \notag \\
    &= \frac{e^{-i \frac{k}{2j}}}{2i} \left[ e^{i \frac{k}{j} J_z} J_x - e^{ -i \frac{k}{j} J_z} J_x + i e^{i \frac{k}{j} J_z} J_y + i e^{ -i \frac{k}{j} J_z} J_y \right] \notag \\
    &= e^{-i \frac{k}{2j}} \left[ \cos\left(\frac{k}{j} J_z\right) J_y + \sin\left(\frac{k}{j} J_z\right) J_x \right].
\end{align}
The last transformation is straightforward as $J_z$ commutes with $J_z^2$:
\begin{align}\label{Eq:nonlinear:Jz}
    e^{i \frac{k}{2j}J_z^2} J_z e^{-i \frac{k}{2j}J_z^2}  = J_z.
\end{align}

\textit{Linear part:} To understand the action of a linear operator $\exp\left(i p J_y\right)$ in the $J_z$ eigenstates, we use Baker--Campbell--Hausdorff formula:
\begin{align}
    {[J_y, J_x]}_n = [J_y, {[J_y, J_x]}_{n-1}] \quad \text{with} \quad {[J_y, J_x]}_0 = J_x.
\end{align}
Thus, the transformation of $J_x$ under the linear part is given by
\begin{align}\label{Eq:linear:Jx}
    e^{i p J_y} J_x e^{-i p J_y} &= \sum_{n=0}^{\infty} \frac{{(i p)}^n}{n!}{[J_y, J_x]}_n \notag \\
    &= J_x + {(i p)} {[J_y, J_x]} + \frac{{(i p)}^2}{2!} [J_y, {[J_y, J_x]}] + \frac{{(i p)}^3}{3!} [J_y, {[J_y, {[J_y, J_x]}]}] + \dots \notag\\
    &= J_x + p J_z - \frac{p^2}{2!} J_x - \frac{p^3}{3!} J_z + \dots \notag \\
    &= J_x \cos(p) + J_z \sin(p). 
\end{align}
Since the operator $J_y$ commutes with itself, its transformation is trivial:
\begin{align}\label{Eq:linear:Jy}
    e^{i p J_y} J_y e^{-i p J_y} = J_y.
\end{align}
Following the method used for $J_x$, the transformation of $J_z$ under the linear part is given as follows:
\begin{align}\label{Eq:linear:Jz}
    e^{i p J_y} J_z e^{-i p J_y} &= \sum_{n=0}^{\infty} \frac{{(i p)}^n}{n!}{[J_y, J_z]}_n \notag\\
    &= J_z + {(i p)} {[J_y, J_z]} + \frac{{(i p)}^2}{2!} [J_y, {[J_y, J_z]}] + \frac{{(i p)}^3}{3!} [J_y, {[J_y, {[J_y, J_z]}]}] + \dots \notag\\
    &= J_z - p J_x - \frac{p^2}{2!} J_z + \frac{p^3}{3!} J_x + \dots \notag\\
    &= -J_x \sin(p) + J_z \cos(p).
\end{align}

\textit{Complete transformation:} The nonlinear part [Eqs.~\eqref{Eq:nonlinear:Jx},~\eqref{Eq:nonlinear:Jy}, and~\eqref{Eq:nonlinear:Jz}] and linear part [Eqs.~\eqref{Eq:linear:Jx},~\eqref{Eq:linear:Jy}, and~\eqref{Eq:linear:Jz}] provide all the ingredients required to find the full transformation. In order to combine them, we can use the following identity (see supplementary material of~\cite{purohit2025double}):
\begin{align}
    e^{i g(\mathbf{J})} f\left(\mathbf{J}\right) e^{-i g(\mathbf{J})} = f\left(e^{i g(\mathbf{J})} \mathbf{J} e^{-i g(\mathbf{J})}\right),
\end{align}
where $f(\mathbf{J})$ is an arbitrary polynomial function and $g(\mathbf{J})$ is an arbitrary function of the angular momentum operators $\mathbf{J} = (J_x, J_y, J_z)$. Using this identity, we get the transformation of $J_x$ under the complete Floquet operator as follows:
\begin{align}\label{Eq:complete:Jx}
    \mathcal{U}^\dagger J_x \mathcal{U} &= e^{i p J_y} \left[ e^{i \frac{k}{2j}J_z^2} J_x e^{-i \frac{k}{2j}J_z^2} \right] e^{-i p J_y} \notag \\
    &= e^{-i \frac{k}{2j}} \; e^{i p J_y} \left( \cos\left(\frac{k}{j} J_z\right) J_x - \sin\left(\frac{k}{j} J_z\right) J_y \right) e^{-i p J_y} \notag \\
    &= e^{-i \frac{k}{2j}} \left\lbrace \cos\left[\frac{k}{j} \left(-J_x \sin(p) + J_z \cos(p)\right)\right] \left(J_x \cos(p) + J_z \sin(p)\right) \right. \notag \notag \\ 
    & \qquad \qquad - \left. \sin\left[\frac{k}{j} \left(-J_x \sin(p) + J_z \cos(p)\right)\right] J_y \right\rbrace.
\end{align}
Similarly, the transformation of $J_y$ under the complete Floquet operator is given by
\begin{align}\label{Eq:complete:Jy}
    \mathcal{U}^\dagger J_y \mathcal{U} &= e^{i p J_y} \left[ e^{i \frac{k}{2j}J_z^2} J_y e^{-i \frac{k}{2j}J_z^2} \right] e^{-i p J_y} \notag \\
    &= e^{-i \frac{k}{2j}} \; e^{i p J_y} \left( \cos\left(\frac{k}{j} J_z\right) J_y + \sin\left(\frac{k}{j} J_z\right) J_x \right) e^{-i p J_y} \notag \\
    &= e^{-i \frac{k}{2j}} \left\lbrace \cos\left[\frac{k}{j} \left(-J_x \sin(p) + J_z \cos(p)\right)\right] J_y \right. \notag \\ 
    & \qquad \qquad + \left. \sin\left[\frac{k}{j} \left(-J_x \sin(p) + J_z \cos(p)\right)\right] \left(J_x \cos(p) + J_z \sin(p)\right) \right\rbrace.
\end{align}
Finally, the transformation of $J_z$ under the complete Floquet operator is given by
\begin{align}\label{Eq:complete:Jz}
    \mathcal{U}^\dagger J_z \mathcal{U} &= e^{i p J_y} \left[ e^{i \frac{k}{2j}J_z^2} J_z e^{-i \frac{k}{2j}J_z^2} \right] e^{-i p J_y} \notag \\
    &= -J_x \sin(p) + J_z \cos(p).
\end{align}
Now, taking the classical limit $\mathbf{X} = (X, Y, Z) = \lim_{j\to\infty} \frac{\mathbf{J}}{j}$, we get the classical map $\mathbf{X}' = \mathbf{F}(\mathbf{X})$ as follows:
\begin{align}\label{Eq:map}
    X' &= \cos \left[k(-X \sin p + Z \cos p)\right] (X \cos p + Z \sin p) - \sin \left[k(-X \sin p + Z \cos p)\right] Y, \notag \\
    Y' &= \cos \left[k(-X \sin p + Z \cos p)\right] Y + \sin \left[k(-X \sin p + Z \cos p)\right] (X \cos p + Z \sin p), \notag \\
    Z' &= -X \sin p + Z \cos p,
\end{align}
with constraint $X^2 + Y^2 + Z^2 = 1$. The classical dynamics of the kicked top are determined by two parameters: the kicking strength $k$ and the precession angle $p$. For a particular point $\mathbf{X}_n$ on the phase space, the above classical map gives the next point $\mathbf{X}_{n+1}$. The trajectory of the kicked top is then obtained by iterating this map over discrete time $n \in \mathds{Z}$.

\end{document}